\documentclass[3p,authoryear,review]{elsarticle}
\journal{Omega}

\usepackage[colorlinks,pdfauthor=author]{hyperref}
\usepackage{amsmath}
\usepackage{amsfonts}
\usepackage{amsthm}
\usepackage{bm}
\usepackage{accents}
\usepackage{booktabs}
\usepackage{tabularx}
\usepackage[title]{appendix}

\newtheorem{prop}{Proposition}

\makeatletter
\DeclareRobustCommand{\rvdots}{%
\vbox{
\baselineskip4\p@\lineskiplimit\z@
\kern-\p@
\hbox{.}\hbox{.}\hbox{.}
}}
\makeatother

\newcolumntype{R}{>{\raggedleft\arraybackslash}X}
 
\usepackage{regexpatch}
\makeatletter
\regexpatchcmd\ps@pprintTitle
{\cE\}\ \cE\}}{\cE\}\cE\}}
{}{\FailedToPatch}
\makeatother

\begin{document}

\begin{frontmatter}

\title{{Flexible global forecast combinations}}
\author[add1,add2]{Ryan Thompson}
\author[add3]{Yilin Qian}
\author[add3]{Andrey L. Vasnev\corref{cor}}
\ead{andrey.vasnev@sydney.edu.au}
\cortext[cor]{Corresponding author.}
\address[add1]{School of Mathematics and Statistics, University of New South Wales, NSW 2052 Australia}
\address[add2]{Data61, Commonwealth Scientific and Industrial Research Organisation, NSW 2015 Australia}
\address[add3]{Discipline of Business Analytics, The University of Sydney, NSW 2006 Australia}

\begin{abstract}
Forecast combination---the aggregation of individual forecasts from multiple experts or models---is a proven approach to economic forecasting. To date, research on economic forecasting has concentrated on local combination methods, which handle separate but related forecasting tasks in isolation. Yet, it has been known for over two decades in the machine learning community that global methods, which exploit task-relatedness, can improve on local methods that ignore it. Motivated by the possibility for improvement, this paper introduces a framework for globally combining forecasts while being flexible to the level of task-relatedness. Through our framework, we develop global versions of several existing forecast combinations. To evaluate the efficacy of these new global forecast combinations, we conduct extensive comparisons using synthetic and real data. Our real data comparisons, which involve forecasts of core economic indicators in the Eurozone, provide empirical evidence that the accuracy of global combinations of economic forecasts can surpass local combinations.
\end{abstract}


\end{frontmatter}

\section{Introduction}
\label{sec:introduction}

Forecast combinations---aggregations of multiple individual forecasts---are one of the most persistently reported empirical successes in forecasting. As a key economic institution, the European Central Bank elicits economic forecasts every quarter for the Eurozone from more than one hundred forecasters, an exercise known as the Survey of Professional Forecasters (SPF). Each forecaster has unique expertise, and some possess private information, so combining is a means to a more accurate and robust projection of the economy than any one forecaster could alone produce. For this reason, the Federal Reserve Bank of Philadelphia runs a similar survey by the same name for the United States. Exactly how to combine forecasts from these surveys is a long-standing problem.

\citet{Bates1969} and later \cite{Newbold1974} and \citet{Granger1984} linearly combined forecasts using variance-minimising weights constrained to sum to one---so-called optimal weights. When the forecasts are unbiased, these weights are optimal in a mean square error sense. In practice, however, they are often beaten by equal weights, a curious phenomenon \citet{Stock2004} called the `forecast combination puzzle'. This puzzle, explained theoretically by \citet{Claeskens2016}, has spurred a formidable research effort to devise improved weighting schemes. \citet{Hansen2008} studied weights that minimise Mallow's criterion, which adds a penalty for complexity. To guarantee a convex combination, \citet{Conflitti2015} added a restriction to prevent negative weights. \citet{Matsypura2018} performed a combinatorial search for the best subset of forecasts to equally weight, and a similar scheme was proposed in \citet{Diebold2019} using an $l_1$-norm penalty. To handle highly correlated forecasts, \citet{Radchenko2023} allowed negative weights but subjected them to a trimming threshold. For other examples in this line of work, see \citet{Yang2004}, \citet{Aiolfi2006}, \citet{Capistran2009}, \citet{Poncela2011}, \citet{Genre2013}, \citet{Burgi2017}, \citet{Kourentzes2019}, and \citet{Qian2022}.

Common to all of the above papers is a focus on using local information to fit the weights, i.e., information that only concerns the forecast target. When just one variable needs forecasting, this approach is sensible. However, it is rare in economics to forecast only a single variable. Instead, forecasts of multiple variables are needed to paint a detailed picture of the economy, core examples being growth, inflation, and unemployment. In addition, policymakers often require forecasts of the economy at different time horizons to facilitate planning. The European Central Bank SPF indeed captures forecasts of multiple variables at multiple horizons, and each variable-horizon pair constitutes an individual forecasting task. Yet, these tasks are not independent; rather, they are highly related. For instance, Okun's law stipulates a strong negative correlation between growth and unemployment \citep{Okun1962}. The Phillips curve sets forth a similar relationship between unemployment and inflation \citep{Phillips1958}. It is not unrealistic to expect then that a forecaster's competence in predicting one variable might contain some signal about their competence in predicting another. This possibility motivates us to consider forecast combinations derived from global information shared across related tasks.

The idea of sharing information between prediction tasks emerged during the 1990s in the machine learning community, where it is known as multi-task learning \citep{Caruana1997}. A vast literature now exists on multi-task learning owing to its success; the interested reader is referred to \citet{Zhang2022} for a comprehensive survey. Research in the forecasting community itself has lately trended towards multi-task learning \citep{Laptev2017,Salinas2020,Godahewa2021,Montero-Manso2021}. In the 2018 M4 Competition, where time series were drawn independently from a large pool, global methods that shared information across forecasting tasks took out the top-three places \citep{Makridakis2020}. Of these three, the second-place method by \citet{Montero-Manso2020} bears some relation to this work. Their method combined forecasts from a handful of classic time series models using weights from gradient boosted trees. The trees were grown on thousands of time series, enabling weights to be learned across tasks. Though similar, their problem is distinct from the economic forecast combination problem that is the main focus of this paper. Whereas \citet{Montero-Manso2020} combined a small number of forecasts for a large number of tasks drawn independently from a large pool, we combine a large number of forecasts for a small number of related tasks. Elaborate approaches involving boosted trees are not feasible in our setting.

In light of the preceding discussion, this paper proposes a new framework for globally combining forecasts. Our framework minimises a global loss function comprised of individual forecasting tasks. The framework is flexible to the level of relatedness among the different tasks. Specifically, using a task-coupling penalty, we interpolate between fully local combination, where all tasks are heterogeneous, and fully global combination, where all tasks are homogeneous. The best interpolation is determined in a data-driven fashion. Via this framework, we `globalise' the weighting schemes of \citet{Bates1969}, \citet{Conflitti2015}, and \citet{Matsypura2018}. We then evaluate the new global combinations in both simulation and an application to expert forecasts from the European Central Bank SPF.\footnote{The literature on the European Central Bank and Federal Reserve Bank of Philadelphia SPFs typically refers to individual forecasts as `expert forecasts'; see footnotes 6 and 7 in \citet{Magnus2023} for papers that use those surveys. Expert forecasts often include judgement that is now recognised as an important element in forecasting and can be used to adjust individual model output \citep{Lawrence2006} or model selection/combination \citep{Petropoulos2018}. In many areas, `judgemental forecasts' is a more common term; see \citet{Lawrence2006}. Our methodology is also applicable to those areas.} The results indicate neither fully local nor fully global combination uniformly performs best. Instead, combinations that lie somewhere between these extremes typically lead to the best out-of-sample performance. We also show the benefits of our framework on model-based forecasts of economic and financial time series from the M4 Competition in \ref{app:m4}.

The paper is organised into six sections. Section~\ref{sec:methodology} introduces the proposed framework for globally combining forecasts. Section~\ref{sec:optimisation} addresses computation of the new combinations. Section~\ref{sec:synthetic} presents numerical experiments that gauge the benefits of globalisation. Section~\ref{sec:survey} describes empirical comparisons of the new methods in application. Section~\ref{sec:conclusion} closes the paper. Proofs are in \ref{app:proofs}, additional synthetic data experiments in \ref{app:B}, and additional empirical results in \ref{app:C}.

\section{Global forecast combinations}
\label{sec:methodology}

\subsection{Single-task forecast combination}

To set the scene for our framework, we first describe the traditional single-task forecast combination problem. Let $y\in\mathbb{R}$ be the forecast target and $\bm{f}=(f_1,\ldots,f_p)^\top\in\mathbb{R}^p$ be forecasts of $y$. Denote by $\bm{e}=y\bm{1}-\bm{f}$ the forecast errors. It is customary to assume the errors satisfy $\operatorname{E}(\bm{e})=\bm{0}$ and $\operatorname{Var}(\bm{e})=\bm{\Sigma}$, where $\bm{\Sigma}$ is a $p\times p$ positive-definite matrix. Consider the linear combination forecast $\tilde{f}=\bm{f}^\top\bm{w}$, where $\bm{w}=(w_1,\ldots,w_p)^\top\in\mathbb{R}^p$ are unit sum weights controlling the contribution of individual forecasts to the combination forecast.

Since the forecasts are unbiased and the weights sum to one, the mean square error minimising forecast combination is that which minimises the combination forecast error variance $\operatorname{Var}(\bm{e}^\top\bm{w})=\bm{w}^\top\bm{\Sigma}\bm{w}$. This minimisation is performed with respect to a constraint set $\mathcal{W}$:
\begin{equation*}
\underset{\bm{w}\in\mathcal{W}}{\min}\,\bm{w}^\top\bm{\Sigma}\bm{w}.
\end{equation*}
The simplest configuration of the constraint set is $\mathcal{W}^\mathrm{eql}=\{\bm{1}/p\}$, yielding \emph{equal weights}. Using $\mathcal{W}^\mathrm{opt}=\{\bm{w}\in\mathbb{R}^p:\bm{1}^\top\bm{w}=1\}$ leads to \emph{optimal weights} as proposed by \citet{Bates1969}. The constraint set $\mathcal{W}^\mathrm{optcvx}=\{\bm{w}\in\mathbb{R}^p:\bm{1}^\top\bm{w}=1,\bm{w}\geq\bm{0}\}$, as studied by \citet{Conflitti2015}, adds a nonnegativity condition to guarantee a convex combination. The resulting weights are referred to hereafter as \emph{optimal convex weights}. A more elaborate configuration, $\mathcal{W}^\mathrm{opteql}=\{\bm{w}\in\mathbb{R}^p:\bm{1}^\top\bm{w}=1,\bm{w}=\bm{z}/(\bm{1}^\top\bm{z}),\bm{z}\in\{0,1\}^p\}$, produces equal weights restricted to an optimal subset of forecasts. These weights were investigated by \citet{Matsypura2018} and are referred to hereafter as \emph{optimal equal weights}. Here, $\bm{z}$ is a vector of $p$ binary variables $z_j$ ($j=1,\ldots,p$) where $z_j$ assumes the value one if forecast $j$ is selected for inclusion in the combination and zero otherwise. The constraint $\bm{w}=\bm{z}/(\bm{1}^\top\bm{z})$ guarantees the selected forecasts are equally-weighted. Other weighting schemes can also be cast in this setup by appropriately choosing $\mathcal{W}$.

When the covariance matrix $\bm{\Sigma}$ is large-dimensional and estimated from data, it can be helpful to include a shrinkage penalty in the objective function \citep[e.g.,][]{Roccazzella2022}:
\begin{equation}
\label{eq:local1}
\underset{\bm{w}\in\mathcal{W}}{\min}\,\bm{w}^\top\bm{\Sigma}\bm{w}+\lambda\|\bm{w}\|_q^q,\quad q\in\{1,2\},
\end{equation}
where $\lambda\geq0$. Setting $q=2$ yields a ridge penalty \citep{Hoerl1970}, while $q=1$ yields a lasso penalty \citep{Tibshirani1996}. When $q=2$, the objective can be rearranged as $\bm{w}^\top(\bm{\Sigma}+\lambda\bm{I})\bm{w}$, so the ridge penalty has the effect of shrinking the covariance matrix towards the identity matrix $\bm{I}$, thereby stabilising the objective. The lasso penalty has a similar stabilising effect. Though there exist numerous covariance estimators that explicitly perform shrinkage \citep{Ledoit2004,Schafer2005,Touloumis2015}, these do not accommodate missing data. Missing data is an important empirical consideration, discussed further in Section~\ref{sec:survey}. On the other hand, it is straightforward to mimic the effect of shrinkage by plugging a standard missing-data covariance estimator into \eqref{eq:local1}. Under all the aforementioned configurations of $\mathcal{W}$, the limiting shrinkage case ($\lambda\to\infty$) leads to equal weights as the optimal solution when $q=2$.

\subsection{Multi-task forecast combination}

The problem described above concerns one forecasting task $y$. Suppose now we have multiple tasks $\bm{y}=(y^{(1)},\ldots,y^{(m)})^\top\in\mathbb{R}^m$. The $m$ tasks may comprise, e.g., different variables or different forecast horizons. We index all quantities relating to the $k$th component by superscript $(k)$. Hence, the combination forecast $\tilde{\bm{f}}=(\tilde{f}^{(1)},\ldots,\tilde{f}^{(m)})^\top\in\mathbb{R}^m$ has elements $\tilde{f}^{(k)}=\bm{f}^{(k)\top}\bm{w}^{(k)}$, where $\bm{f}^{(k)}=(f_1^{(k)},\ldots,f_p^{(k)})^\top$ and $\bm{w}^{(k)}=(w_1^{(k)},\ldots,w_p^{(k)})^\top$. The errors are $\bm{e}^{(k)}=y^{(k)}\bm{1}-\bm{f}^{(k)}$ with $\operatorname{Var}(\bm{e}^{(k)})=\bm{\Sigma}^{(k)}$.

Though the multi-task setup is typical of economics, research to date has treated the tasks in isolation, using weights fit on a per-task basis:
\begin{equation}
\label{eq:local2}
\underset{\bm{w}^{(1)},\ldots,\bm{w}^{(m)}\in\mathcal{W}}{\min}\,\sum_{k=1}^m\left(\bm{w}^{(k)\top}\bm{\Sigma}^{(k)}\bm{w}^{(k)}+\lambda\|\bm{w}^{(k)}\|_q^q\right).
\end{equation}
This combination is local because the individual tasks are in no way linked, i.e., solving optimisation problem \eqref{eq:local1} for each task individually leads to the same weights as solving optimisation problem \eqref{eq:local2}. Information from one task that might be relevant to other tasks is neglected. Instead, one can consider a single vector of weights that is a minimiser of the total loss across all tasks:
\begin{equation}
\label{eq:hard-global}
\underset{\bm{w}\in\mathcal{W}}{\min}\,\sum_{k=1}^m\left(\bm{w}^\top\bm{\Sigma}^{(k)}\bm{w}+\lambda\|\bm{w}\|_q^q\right).
\end{equation}
This combination is global insofar as the resulting weights take into account information contained in all tasks. Since the loss term in the objective can be expressed equivalently as $\bm{w}^\top(\sum_{k=1}^m\bm{\Sigma}^{(k)})\bm{w}$, this approach can be interpreted as averaging over the task-specific covariance matrices. When the covariance matrices are estimated by the sample covariance matrix, averaging is the same as estimating a single covariance matrix after aggregating data from different tasks. Unfortunately, an implicit assumption underlies this approach that the tasks are completely homogeneous. This assumption might be unreasonably strong in practice and could harm forecast performance.

\begin{figure}[ht!]
\centering
\resizebox{6in}{!}{\includegraphics{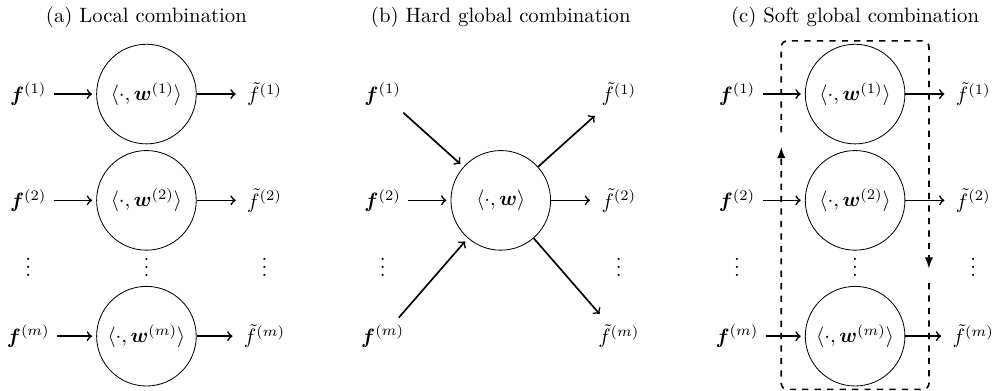}}
\caption{Global and local forecast combination frameworks. The notation $\langle\bm{x},\bm{y}\rangle=\bm{x}^\top\bm{y}$ represents the dot product of two vectors $\bm{x}\in\mathbb{R}^p$ and $\bm{y}\in\mathbb{R}^p$. Local combination learns different weight vectors for each task independently of other tasks. Hard global combination learns one weight vector for all tasks. Soft global combination learns different weight vectors for each task while sharing information between tasks.}
\label{fig:frameworks}
\end{figure}

Rather than committing to a fully local or fully global approach, one can consider bridging the two approaches using per-task weights that are globally regularised:
\begin{equation}
\label{eq:soft-global}
\underset{\substack{\bm{w}^{(1)},\ldots,\bm{w}^{(m)}\in\mathcal{W} \\ \bar{\bm{w}}\in\mathbb{R}^p}}{\min}\,\sum_{k=1}^m\left(\bm{w}^{(k)\top}\bm{\Sigma}^{(k)}\bm{w}^{(k)}+\lambda\|\bm{w}^{(k)}\|_q^q+\gamma\|\bar{\bm{w}}-\bm{w}^{(k)}\|_q^q\right),\quad q\in\{1,2\}.
\end{equation}
Here, the penalty $\gamma\sum_{k=1}^m\|\bar{\bm{w}}-\bm{w}^{(k)}\|_q^q$ with $\gamma\geq0$ is a device to incorporate global information into the per-task weights. It achieves this goal by penalising departures from an auxiliary weight vector $\bar{\bm{w}}$ common to all tasks, where the departures are measured as squared deviations ($q=2$) or absolute deviations ($q=1$). Regardless of $q$, taking $\gamma\to\infty$ yields global combination \eqref{eq:hard-global}, while taking $\gamma\to0$ yields local combination \eqref{eq:local2}. Hereafter, we refer to the limiting case $\gamma\to\infty$ as `hard' global combination, and the case with finite nonzero values of $\gamma$ as `soft' global combination. These different cases are depicted in Figure~\ref{fig:frameworks}.
The value of $\gamma$ should reflect the level of relatedness among tasks---larger values encourage homogeneity, while smaller values promote heterogeneity. The best value in terms of out-of-sample forecast performance is usually unknown in application but is estimable from data.

\subsection{Alternative formulations}

The optimisation problem \eqref{eq:soft-global} can be cast solely in terms of the per-task weights $\bm{w}^{(1)},\ldots,\bm{w}^{(m)}$:
\begin{equation*}
\underset{\bm{w}^{(1)},\ldots,\bm{w}^{(m)}\in\mathcal{W}}{\min}\,\sum_{k=1}^m\left(\bm{w}^{(k)\top}\bm{\Sigma}^{(k)}\bm{w}^{(k)}+\lambda\|\bm{w}^{(k)}\|_q^q\right)+\Omega_{\gamma,q}(\bm{w}^{(1)},\ldots,\bm{w}^{(m)}),
\end{equation*}
where
\begin{equation}
\label{eq:reg}
\Omega_{\gamma,q}(\bm{w}^{(1)},\ldots,\bm{w}^{(m)})=\underset{\bar{\bm{w}}\in\mathbb{R}^p}{\min}\,\gamma\sum_{k=1}^m\|\bar{\bm{w}}-\bm{w}^{(k)}\|_q^q,\quad q\in\{1,2\}.
\end{equation}
Penalties like $\Omega_{\gamma,q}$, which penalise departures from a common parameter vector, first appeared in the context of multi-task kernel learning \citep{Evgeniou2004,Evgeniou2005}. 

When the departures are measured as squared deviations (i.e., $q=2$), it is not difficult to obtain an analytical solution:
\begin{equation*}
\Omega_{\gamma,2}(\bm{w}^{(1)},\ldots,\bm{w}^{(m)})=\gamma\sum_{k=1}^m\left\|\frac{1}{m}\sum_{l=1}^m\bm{w}^{(l)}-\bm{w}^{(k)}\right\|_2^2.
\end{equation*}
That is, the optimal value of the common parameter vector $\bar{\bm{w}}$ is the average of the individual parameter vectors $\bm{w}^{(1)},\ldots,\bm{w}^{(m)}$. One can thus interpret our approach as finding per-task weights within a certain distance of the average weight vector. Some additional algebra gives an alternative expression for $\Omega_{\gamma,2}$:
\begin{equation*}
\Omega_{\gamma,2}(\bm{w}^{(1)},\ldots,\bm{w}^{(m)})=\frac{\gamma}{m}\sum_{k=1}^m\sum_{l=1}^k\|\bm{w}^{(l)}-\bm{w}^{(k)}\|_2^2.
\end{equation*}
This expression highlights that our approach explicitly penalises mutual distances between local weight vectors. Our experience is that formulating soft global combination using either of the above closed-form solutions yield computational performance similar to that of \eqref{eq:soft-global}, provided the number of tasks $m$ is not large. When $m$ is large, these solutions involve many more quadratic terms in the objective, which can impede computation. For instance, under the simulation design of Section~\ref{sec:synthetic} when $m=10$ and $p=50$, it takes roughly six times longer to solve for optimal weights when using the second of the above closed-form solutions.

\begin{prop}
\label{prop:extra}
When $q=2$, the optimisation problem \eqref{eq:soft-global} can also be expressed as
\begin{equation}
\label{eq:min_wk}
\underset{\substack{\bm{w}^{(1)},\ldots,\bm{w}^{(m)}\in\mathcal{W} }}{\min}\,\sum_{k=1}^m\left(\bm{w}^{(k)\top}[\bm{\Sigma}^{(k)}+(\lambda+\gamma)\bm{I}]\bm{w}^{(k)}-\gamma\bar{\bm{w}}^{\star\top} \bar{\bm{w}}^\star\right),
\end{equation}
where $\bar{\bm{w}}^\star=m^{-1}\sum_{k=1}^m\bm{w}^{(k)}$ is the optimal value of the common parameter vector.
\begin{proof}
See \ref{app:proof_extra}.
\end{proof}
\end{prop}

Form \eqref{eq:min_wk} reveals that $\gamma$ plays a dual role. It shrinks towards equal weights when it appears in front of $\bm{I}$, similar to $\lambda$, but it also pushes towards a corner solution via the last term. While the full explicit solution cannot be derived for $\gamma\neq0$, it is possible to prove the following proposition.
\begin{prop}
\label{prop:01}
The optimal solution of problem \eqref{eq:min_wk} when $\mathcal{W}=\mathcal{W}^\mathrm{opt}$ satisfies 
\begin{equation*}
\bm{w}^{(k)}=\frac{1-\gamma\bm{1}^\top A^{(k)}BD\bar{\bm{w}}^\star}{\bm{1}^\top A^{(k)} \bm{1}}A^{(k)}\bm{1}+\gamma A^{(k)}BD\bar{\bm{w}}^\star,
\end{equation*}
where $A^{(k)}=[\bm{\Sigma}^{(k)}+(\lambda+\gamma)\bm{I}]^{-1}$, $B=\left(\bm{I}-\gamma/m\sum_{l=1}^{m} A^{(l)}\right)^{-1}$, and $D=\left(\bm{I}+\gamma/m\sum_{l=1}^m A^{(l)}B\right)^{-1}$.
\begin{proof}
See \ref{app:proof}.
\end{proof}
\end{prop}
For $\gamma=0$, we get an explicit solution $\bm{w}^{(k)} = {A^{(k)} \bm{1}}/(\bm{1}^\top A^{(k)} \bm{1})$ which is optimal weights of \citet{Bates1969} shrunk towards equal weights by $\lambda$. When $\gamma\ne 0$, it helps $\lambda$ with shrinkage, as expected, but also enters in a highly nonlinear way via $B$ and $D$, so the total effect of $\gamma$ is difficult to discern. 

\subsection{Task grouping}

Sometimes it can be useful to limit the flow of information between certain tasks, e.g., when one or more tasks are unrelated. For this purpose, denote by $\mathcal{G}:=\{\mathcal{G}_1,\ldots,\mathcal{G}_g\}$ a collection of $g$ groups of tasks, where $\mathcal{G}_l\subseteq\{1,\ldots,m\}$, $\mathcal{G}_1\cup\cdots\cup\mathcal{G}_g=\{1,\ldots,m\}$, and $\mathcal{G}_l\cap\mathcal{G}_k=\emptyset$ for all $l\neq k$. Using this notation, one can modify $\Omega_{\gamma,q}$ to impose the restriction that only tasks within the same group share information:
\begin{equation*}
\Omega_{\gamma,q}(\bm{w}^{(1)},\ldots,\bm{w}^{(m)})=\underset{\bar{\bm{w}}^{(1)},\ldots,\bar{\bm{w}}^{(g)}\in\mathbb{R}^p}{\min}\,\gamma\sum_{l=1}^g\sum_{k\in\mathcal{G}_l}\|\bar{\bm{w}}^{(l)}-\bm{w}^{(k)}\|_q^q,\quad q\in\{1,2\},
\end{equation*}
where $\bar{\bm{w}}^{(l)}$ is an auxiliary weight vector for the $l$th group. When $\mathcal{G}$ consists of just one group, this grouped version of the penalty reduces to \eqref{eq:reg}. Conversely, when $\mathcal{G}$ consists of $m$ groups, the grouped penalty has no globalisation effect, i.e., it leads to local combination. The grouped version is helpful in our application to the SPF data in Section~\ref{sec:survey} where we study different groups of variables and forecast horizons.

\subsection{Task scaling}

If the tasks under consideration vary in difficulty, one or more tasks might dominate the loss component of the objective function. To prevent this behaviour, we consider a scaled version of global combination:
\begin{equation*}
\underset{\bm{w}^{(1)},\ldots,\bm{w}^{(m)}\in\mathcal{W}}{\min}\,\sum_{k=1}^m\frac{\bm{w}^{(k)\top}\bm{\Sigma}^{(k)}\bm{w}^{(k)}+\lambda\|\bm{w}^{(k)}\|_q^q}{\tau_q^{(k)}}+\Omega_{\gamma,q}(\bm{w}^{(1)},\ldots,\bm{w}^{(m)}),
\end{equation*}
where $\tau_q^{(1)},\ldots,\tau_q^{(m)}>0$ are fixed scaling parameters. If the tasks are to be evenly balanced, a suitable value of $\tau_q^{(k)}$ is the optimal objective value from local combination:
\begin{equation*}
\tau_q^{(k)}=\underset{\bm{w}\in\mathcal{W}}{\min}\,\bm{w}^\top\bm{\Sigma}^{(k)}\bm{w}+\lambda\|\bm{w}\|_q^q.
\end{equation*}
This configuration of $\tau_q^{(k)}$ places all tasks on equal footing, and we use it in all subsequent experiments.

\section{Optimisation}
\label{sec:optimisation}

\subsection{Optimal (convex) weights}

Computation of forecast combinations in our framework varies in complexity according to the weighting scheme, i.e., the specific configuration of $\mathcal{W}$. We begin by describing methods for computation for optimal weights of \citet{Bates1969} and optimal convex weights of \citet{Conflitti2015}, both natural candidates for our framework. The constraint sets $\mathcal{W}^\mathrm{opt}=\{\bm{w}\in\mathbb{R}^p:\bm{1}^\top\bm{w}=1\}$ and $\mathcal{W}^\mathrm{optcvx}=\{\bm{w}\in\mathbb{R}^p:\bm{1}^\top\bm{w}=1,\bm{w}\geq\bm{0}\}$ defining these combinations are convex. All the objective functions described in Section~\ref{sec:methodology} are convex. The resulting convex optimisation problems are efficiently solvable using most mathematical programming solvers; we use \texttt{Gurobi} \citep{gurobi2023}.

\subsection{Optimal equal weights}

Optimal equal weights of \citet{Matsypura2018} are another natural candidate for our framework. The constraint set defining these weights is less tractable than that for optimal weights or optimal convex weights. Recall the set is defined by a mix of continuous and discrete variables:
\begin{equation}
\label{eq:select0}
\mathcal{W}^\mathrm{opteql}=\{\bm{w}\in\mathbb{R}^p:\bm{1}^\top\bm{w}=1,\bm{w}=\bm{z}/(\bm{1}^\top\bm{z}),\bm{z}\in\{0,1\}^p\}.
\end{equation}
The integrality constraint $\bm{z}\in\{0,1\}^p$ is nonconvex but is amenable to a mixed-integer programming solver such as \texttt{Gurobi}. The constraint $\bm{w}=\bm{z}/(\bm{1}^\top\bm{z})$ is also nonconvex but cannot be handled directly by \texttt{Gurobi}. \citet{Matsypura2018} used the decomposition $\mathcal{W}^\mathrm{opteql}=\cup_{s=1,\ldots,p}\mathcal{W}_s^\mathrm{opteql}$, where $\mathcal{W}_s^\mathrm{opteql}=\{\bm{w}\in\mathbb{R}^p:\bm{1}^\top\bm{w}=1,\bm{w}=\bm{z}/s,\bm{z}\in\{0,1\}^p\}$ is the set of all vectors that equally weight $s$ forecasts. Since $s$ is fixed for $\mathcal{W}_s^\mathrm{opteql}$, the constraint $\bm{w}=\bm{z}/s$ is linear. The authors sequentially optimise over $\mathcal{W}_1^\mathrm{opteql},\ldots,\mathcal{W}_p^\mathrm{opteql}$ and retain a solution with minimal objective value. This decomposition approach is, however, infeasible in our framework, because different tasks need not combine the same number of forecasts. To this end, we use a new one-step approach which directly optimises over $\mathcal{W}^\mathrm{opteql}$. Though this new approach is proposed for the purpose of globally combining forecasts, it may be of independent interest for local forecast combination. We have found it to be to be uniformly faster than the approach in \citet{Matsypura2018} in the single-task setting, sometimes by an order of magnitude.

First, we rewrite the constraint $\bm{w}=\bm{z}/(\bm{1}^\top\bm{z})$ as the pair of constraints $\bm{w}s=\bm{z}$ and $s=\bm{1}^\top\bm{z}$, where $s\in\{1,\ldots,p\}$. The new constraint $\bm{w}s=\bm{z}$ is \emph{bilinear} in $\bm{w}$ and $s$, meaning it is linear for fixed $\bm{w}$ or fixed $s$. Though this bilinear constraint remains nonconvex, it is amenable to spatial branch-and-bound techniques \citep{Liberti2008} which are similar to classic branch-and-bound techniques used for handling integrality constraints. As of version 9, released in 2020, \texttt{Gurobi} can solve optimisation problems with bilinear constraints to global optimality. We now rewrite the constraint set \eqref{eq:select0} using the new bilinear constraint representation:
\begin{equation*}
\begin{split}
\mathcal{W}^\mathrm{opteql}=\{\bm{w}\in\mathbb{R}^p:\bm{1}^\top\bm{w}=1,\bm{w}s=\bm{z},s=\bm{1}^\top\bm{z},s\in\{1,\ldots,p\},\bm{z}\in\{0,1\}^p\}.
\end{split}
\end{equation*}
The constraint $s=\bm{1}^\top\bm{z}$ is, in fact, redundant in the above characterisation of $\mathcal{W}^\mathrm{opteql}$ since it is implied by the remaining constraints. Our experience is that \texttt{Gurobi} benefits from excluding it.

\section{Synthetic data experiments}
\label{sec:synthetic}

\subsection{Simulation design}

We evaluate the possible gains from global forecast combination in simulation. We work directly with the forecast errors which are sampled from a $p$-dimensional Gaussian $\bm{e}_t^{(k)}\sim N(\bm{0},\bm{\Sigma}^{(k)})$ for $t=1,\ldots,T$ and $k=1,\ldots,m$. We fix $p=T=50$, so the number of forecasters is of the same order as the number of samples. Different sample sizes are considered in \ref{app:larger}, though the main findings are robust to sample size. The number of tasks $m\in\{2,5,10\}$. The covariance matrices $\bm{\Sigma}^{(1)},\ldots,\bm{\Sigma}^{(m)}$ are constructed element-wise as $\Sigma_{ij}^{(k)}=\sigma_i^{(k)}\sigma_j^{(k)}\rho^{|i-j|}$. The correlation parameter $\rho=0.75$ to induce high correlations between forecasters, typical of forecaster surveys. For forecaster $j=1,\ldots,p$, the standard deviations $\sigma_j^{(k)}$ are generated by drawing random variables uniformly distributed on $[a,b]$ and correlating them with correlation coefficient $\alpha\in\{0,1/3,2/3,1\}$.
The parameter $\alpha$ dictates the level of task-relatedness. As $\alpha$ approaches one, a forecaster's performance on one task is strongly indicative of their performance on other tasks. The converse is true as $\alpha$ approaches zero---a forecaster's performance on one task is weakly indicative of their performance on other tasks. The bounds $a=1$ and $b=3$ so the accuracy of the worst forecaster is up to three times poorer than that of the best forecaster. A visualisation of data from this simulation design is given in \ref{app:visualisation}.

As a measure of out-of-sample accuracy, we report the mean square forecast error on an infinitely large testing set relative to that from an oracle:
\begin{equation*}
\text{MSFE relative to oracle}:=\frac{(\hat{\bm{w}}^{(1)}-\bm{w}^{(1)})^\top\bm{\Sigma}^{(1)}(\hat{\bm{w}}^{(1)}-\bm{w}^{(1)})}{\bm{w}^{(1)\top}\bm{\Sigma}^{(1)}\bm{w}^{(1)}},
\end{equation*}
where $\hat{\bm{w}}^{(1)}$ denotes estimated weights for task one fit using an estimate $\hat{\bm{\Sigma}}^{(1)}$ of the true covariance matrix $\bm{\Sigma}^{(1)}$, and $\bm{w}^{(1)}$ denotes oracle weights fit using $\bm{\Sigma}^{(1)}$. We restrict our attention to the relative forecast error of the first task only to measure the marginal effect of adding additional tasks. The covariance matrices are estimated using the sample covariances $\hat{\Sigma}_{ij}^{(k)}=T^{-1}\sum_{t=1}^Te_{it}^{(k)}e_{jt}^{(k)}$ for all $(i,j)\in\{1,\ldots,p\}^2$.

The shrinkage parameter $\lambda$ is swept over a grid of ten values evenly spaced on a logarithmic scale between $0.001$ and $1000$. For every value of $\lambda$, the globalisation parameter $\gamma$ of soft global combination is swept over the same grid. The best values of $\lambda$ and $\gamma$ are chosen on a validation set constructed independently and identically to the training set, which we remark approximates the precision of leave-one-out cross-validation.

The simulations are run parallel in \texttt{R} \citep{R2023} with \texttt{Gurobi} given a single core of an AMD Ryzen Threadripper 3970x and a 300 second time limit for each value of $\gamma$ and $\lambda$.

\subsection{Forecast performance}

\begin{figure}[ht!]
\centering
\includegraphics{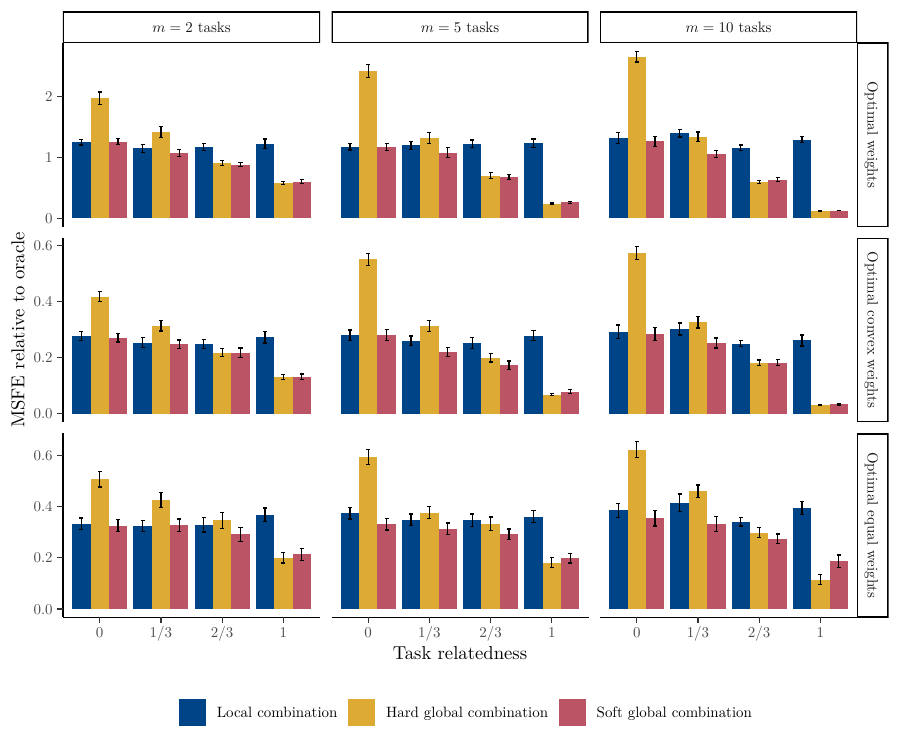}
\caption{Mean square forecast error as a function of task-relatedness parameter $\alpha$ for 30 synthetic datasets with $p=50$ forecasters and $T=50$ samples. Vertical bars represent averages and error bars denote one standard errors. All values are relative to oracle weights.}
\label{fig:simulation-50}
\end{figure}

Figure~\ref{fig:simulation-50} reports the relative forecast errors from 30 simulations.
The first row of plots is where the estimate $\hat{\bm{w}}$ and oracle $\bm{w}$ are fit under the sum to one constraint that defines optimal weights. The second and third rows correspond to the cases where $\hat{\bm{w}}$ and $\bm{w}$ are fit under the constraints that define optimal convex weights and optimal equal weights, respectively. The relative forecast error reported is not comparable across these three weighting schemes since the oracle is different in each case. Our goal is not to compare weighting schemes but rather to measure the benefits of globalisation. The interested reader is referred to \ref{app:equal} for forecasts errors reported relative to equal weights---all key findings below remain the same.

Since local combination ignores information in additional tasks, its performance stays fixed as both the number of tasks and task-relatedness increase. In contrast, the relative forecast error of hard global combination decreases roughly linearly with task-relatedness, providing for substantial improvements when task-relatedness is high. Yet, when task-relatedness is low, hard global combination can underperform relative to local combination. This poor performance is made worse by adding additional tasks.

Soft global combination ameliorates the poor performance of hard global combination when the tasks are unrelated and nearly performs as well as hard global combination when the tasks are identical. There is, of course, a statistical cost to estimating the best level of globalisation. Between the extremes, soft global combination successfully adapts to the level of task-relatedness to improve over both local and hard global combination. The greater the number of tasks, the greater the possibility for improvement.

Among the three weighting schemes, optimal weights benefit most from globalisation. The constraint set that defines optimal weights is unbounded, and thus its relative forecast error can be arbitrarily bad. Optimal convex weights and optimal equal weights are defined by bounded constraint sets, so there exist finite upper bounds on their relative forecast errors. Thus, the opportunity to improve these weights is somewhat less than for optimal weights, yet often still substantial.

In \ref{app:larger}, we provide additional results and extended discussion for $T\in\{25,100,150\}$. For shorter series ($T=25$), soft global combination performs well even when the tasks are unrelated and significantly improves when they are related. Even though the justification is different for longer series, the conclusion is the same: soft global combination is preferred. It gets the best of both worlds regardless of whether the tasks are related.

The soft global combination results in this section correspond to the globalisation penalty configured with squared deviations ($q=2$). Further comparisons in \ref{app:penalty} indicate no material improvement by the absolute deviation penalty ($q=1$), so we restrict our attention to squared deviations hereafter.

\subsection{Recommendations}

The findings from these experiments suggest several recommendations for practitioners. First, consider globalising any forecast combinations when tackling multiple forecasting tasks. The potential gains from globalisation can be significant, even for moderate levels of task-relatedness. Second, unless domain knowledge indicates the tasks are unrelated or strongly related, use soft global combination with cross-validation. Soft global combination with $\gamma$ cross-validated is reasonably robust to task-relatedness, while the downside of applying hard global (local) combination to unrelated (strongly related) tasks is large. Last, when using optimal weights, employ global combination whenever possible since that weighting scheme benefits most from globalisation. The benefits persist even when globalising in tandem with shrinkage.

\section{Survey of Professional Forecasters}
\label{sec:survey}

\subsection{Data and methodology}

The European Central Bank SPF is an ongoing survey eliciting predictions for rates of growth, inflation, and unemployment from forecasters for the Eurozone. The survey has been conducted quarterly since 1999 Q1. In each round, the survey participants are asked to provide predictions of the three variables at several time horizons. We focus on the two rolling horizons in this paper, which are one and two years ahead of the latest available observation of the respective variable. For instance, in the 1999 Q1 survey, one-year forecasts corresponded to 1999 Q3 for growth, December 1999 for inflation, and November 1999 for unemployment.\footnote{To simplify exposition, forecasts of inflation and unemployment are referred to by the quarter they belong to, e.g., December 1999 inflation and November 1999 unemployment are called forecasts of 1999 Q4.} The total number of forecasting tasks $m=6$.

The SPF data is publicly available at the European Central Bank Statistical Data Warehouse (SDW). Actual values of inflation and unemployment are also available at the SDW. Actual values of growth are available from Eurostat. We access data at the SDW using the \texttt{R} package \texttt{ecb} \citep{Persson2022}, and data from Eurostat using the \texttt{R} package \texttt{eurostat} \citep{Lahti2017}. The data used in this paper was retrieved on 17 April 2022. After merging the forecasts and actual values, between $T=85$ and $T=90$ observations are available. The first observations are 1999 Q3 (one-year growth), 1999 Q4 (one-year inflation and unemployment), 2000 Q3 (two-year growth), and 2000 Q4 (two-year inflation and unemployment). The last observation is 2021 Q4.

\begin{figure}[ht!]
\centering
\includegraphics{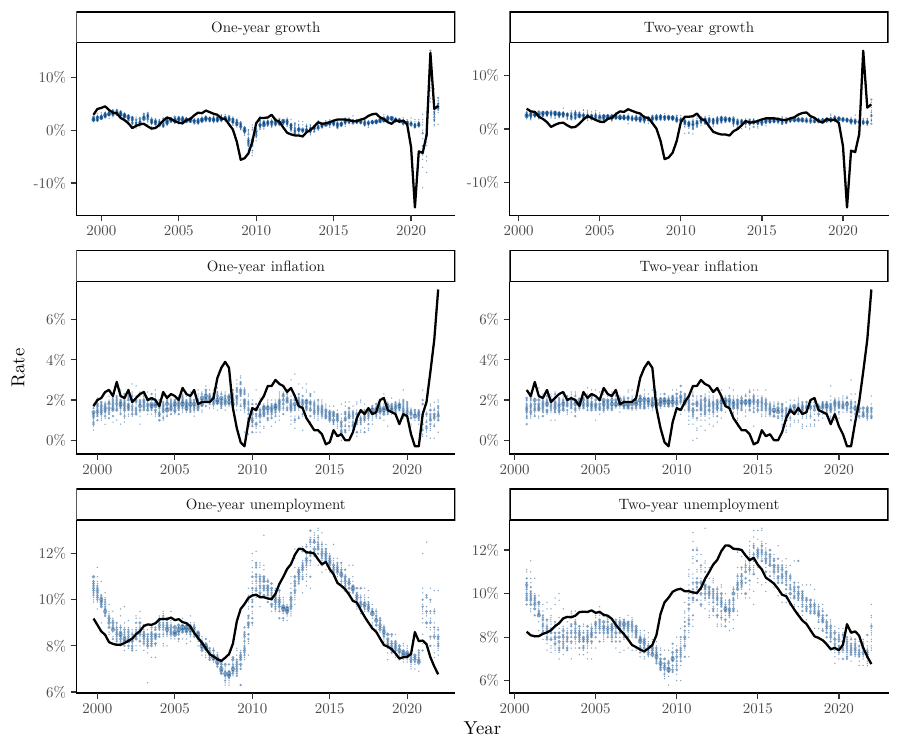}
\caption{Data from the Survey of Professional Forecasters. Points represent forecasts and lines denote actual values of the forecast target. Point sizes reflect the number of equivalent forecasts when rounded to one decimal place.}
\label{fig:spf-data}
\end{figure}

A notable feature of the SPF is that forecasters enter and exit the survey at different times. This aspect of the survey, coupled with periodic nonresponse, gives rise to a sizeable portion of missing data. To deal with this issue, we follow previous works \citep{Matsypura2018,Radchenko2023} and filter the data to only include forecasters who respond for a reasonable number of periods. Specifically, the forecasters who provide a minimum of 40 forecasts (10 years) for every task over the full training set (1999 Q3 to 2019 Q4) are retained. This filtering criterion leads to a dataset comprising $p=34$ forecasters. Figure~\ref{fig:spf-data} plots the filtered forecasts alongside actual values of the forecast targets.

To handle missing values that remain after filtering, the covariance matrices of forecast errors are estimated using all complete pairs of observations: $\hat{\Sigma}_{ij}^{(k)}=|\mathcal{T}_i^{(k)}\cap\mathcal{T}_j^{(k)}|^{-1}\sum_{t\in\mathcal{T}_i^{(k)}\cap\mathcal{T}_j^{(k)}}e_{it}^{(k)}e_{jt}^{(k)}$ for all $(i,j)\in\{1,\ldots,p\}^2$. Here, $\mathcal{T}_i^{(k)}$ denotes the periods in the training set where forecaster $i$ provided a forecast for task $k$. Covariance matrices constructed in this manner are not guaranteed positive-definite. For this reason, we take the positive-definite matrix nearest to $\hat{\bm{\Sigma}}^{(k)}$ using \texttt{nearPD} from the \texttt{R} package \texttt{Matrix} \citep{Bates2022}. The forecast errors are standardised by the standard deviation of the forecast targets as estimated on the training set prior to estimating the covariance matrices.

\subsection{Globalisation path}

The first set of experiments study the evolution of out-of-sample forecast performance as the globalisation parameter $\gamma$ is swept over its support (the `globalisation path'). Here, we take 30 values of $\gamma$ logarithmically spaced between $0.001$ and $1000$. As a measure of out-of-sample accuracy, we report the mean square forecast error on a testing set relative to that from local combination: 
\begin{equation*}
\text{MSFE relative to local}:=\frac{\sum_{t=\underaccent{\bar}{T}-h}^{\bar{T}-h}(y_{t+h}^{(k)}-\tilde{f}_{t+h\mid t}^{(k)(\gamma)})^2}{\sum_{t=\underaccent{\bar}{T}-h}^{\bar{T}-h}(y_{t+h}^{(k)}-\tilde{f}_{t+h\mid t}^{(k)(0)})^2},
\end{equation*}
where, for a given weighting scheme, $\tilde{f}_{t+h\mid t}^{(k)(\gamma)}$ is a global combination forecast of task $k$ at time $t+h$ produced using a training set up to time $t$ with $\gamma\in[0,\infty)$, and $\underaccent{\bar}{T}$ and $\bar{T}$ are the first and last periods in the testing set. The denominator is the mean square forecast error from setting $\gamma=0$, so this metric is the percentage improvement due to globalisation. We pick $\underaccent{\bar}{T}$ and $\bar{T}$ so the testing set is the last five years to 2019 Q4. The period after 2019 Q4, covering the COVID-19 recession and 2021--2022 inflation surge, is considered in separate experiments in Section~\ref{sec:tuned}.

\begin{figure}[ht!]
\centering
\includegraphics{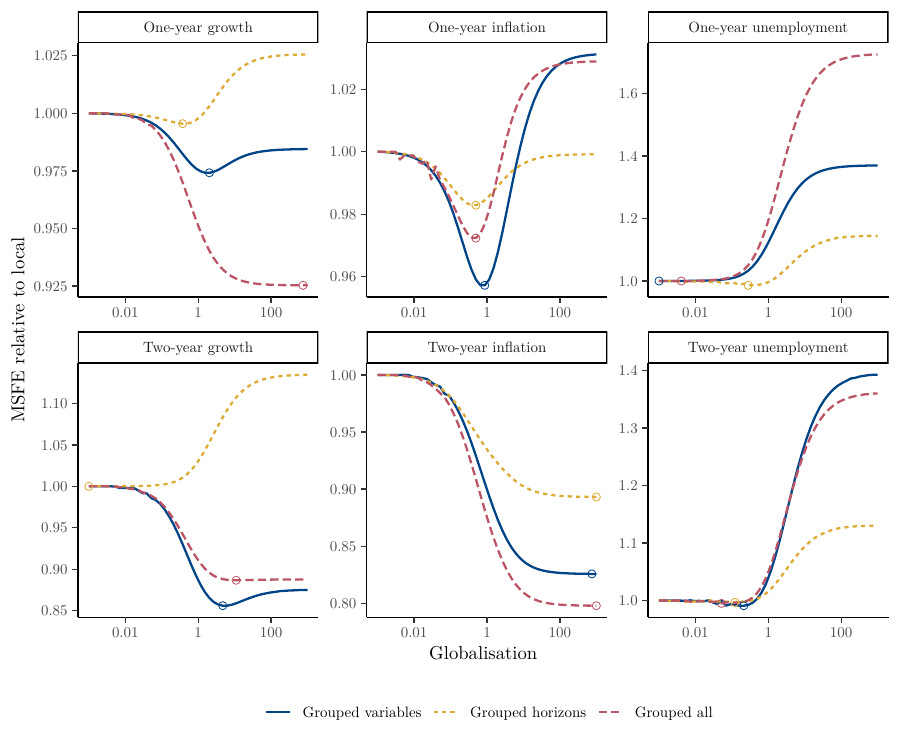}
\caption{Mean square forecast error of \emph{optimal weights} as a function of globalisation parameter $\gamma$ for the Survey of Professional Forecasters. Testing period is 2015 Q1 to 2019 Q4. Minimum of each curve is marked by a circle. All values are relative to local combination ($\gamma\to0$). The shrinkage parameter $\lambda=0.1$. The $x$-axis is in log scale.}
\label{fig:spf-path-optimal}
\end{figure}

\begin{figure}[ht!]
\centering
\includegraphics{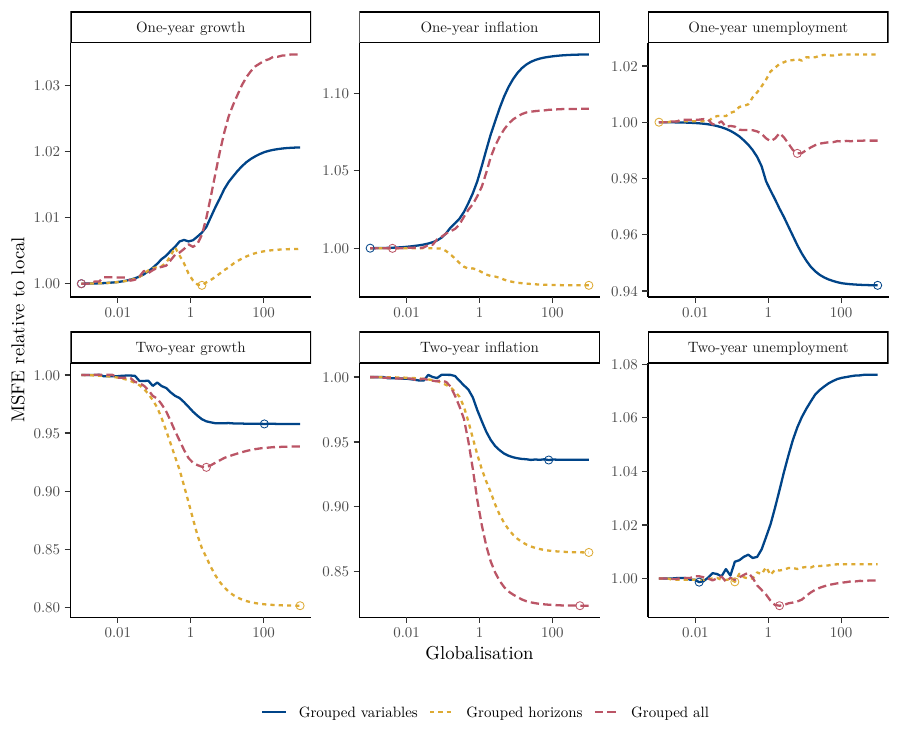}
\caption{Mean square forecast error of \emph{optimal convex weights} as a function of globalisation parameter $\gamma$ for the Survey of Professional Forecasters. Testing period is 2015 Q1 to 2019 Q4. Minimum of each curve is marked by a circle. All values are relative to local combination ($\gamma\to0$). The shrinkage parameter $\lambda=0.1$. The $x$-axis is in log scale.}
\label{fig:spf-path-optimal-convex}
\end{figure}

\begin{figure}[ht!]
\centering
\includegraphics{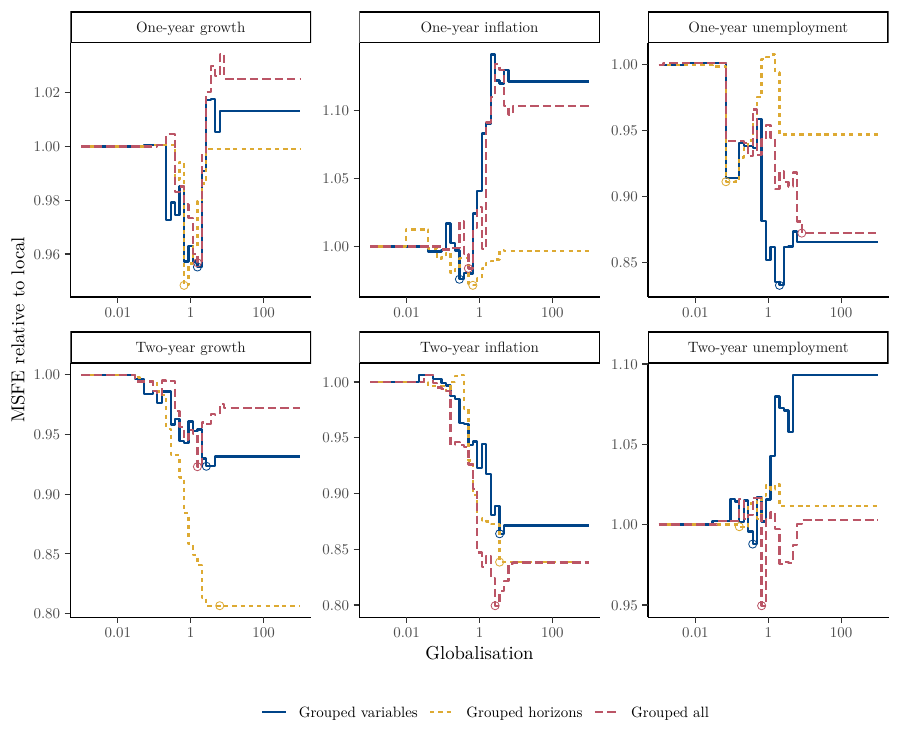}
\caption{Mean square forecast error of \emph{optimal equal weights} as a function of globalisation parameter $\gamma$ for the Survey of Professional Forecasters. Testing period is 2015 Q1 to 2019 Q4. Minimum of each curve is marked by a circle. All values are relative to local combination ($\gamma\to0$). The shrinkage parameter $\lambda=0.1$. The $x$-axis is in log scale.}
\label{fig:spf-path-optimal-equal}
\end{figure}

Figures \ref{fig:spf-path-optimal}, \ref{fig:spf-path-optimal-convex}, and \ref{fig:spf-path-optimal-equal} report the globalisation paths of optimal weights, optimal convex weights, and optimal equal weights for fixed shrinkage parameter $\lambda=0.1$.
The globalisation paths of optimal weights are smooth because the fitted weights are a smooth function of $\gamma$ as Proposition~\ref{prop:01} implies, while those of optimal convex weights and optimal equal weights are nonsmooth. In the case of optimal convex weights, the convexity constraint makes the fitted weights nonsmooth in $\gamma$ when it is binding. The path for optimal equal weights is a step function in $\gamma$ due to the weights being discrete. Three ways of grouping the tasks are considered: grouping variable tasks (group 1: one-year growth, inflation, and unemployment; group 2: two-year growth, inflation, and unemployment); grouping forecast horizon tasks (group 1: one- and two-year growth; group 2: one- and two-year inflation; group 3: one- and two-year unemployment); and grouping all tasks (group 1: one- and two-year growth, inflation, and unemployment). The reader is reminded information flows only between tasks belonging to the same group.

Across all weighting schemes and tasks, there is always a globalisation path that attains its minimum at some positive amount of globalisation. The limiting case $\gamma\to\infty$, hard global combination, is sometimes helpful and sometimes harmful. For instance, growth and inflation realise roughly 15\% improvement from hard global combination (optimal weights, grouped variables) at the two-year horizon while unemployment deteriorates by about 40\% at the same horizon. This behaviour might be attributable to growth and inflation being difficult tasks at the two-year horizon (e.g., expert forecasts of those tasks are not responsive to the COVID effects in 2020 and 2021 as Figure~\ref{fig:spf-data} shows), thus providing a noisy signal to unemployment. However, even in the cases where hard global combination on its own is not useful (such as one- and two-year unemployment forecasts), the optimal choice of $\gamma$ is still positive, and soft global combination can extract benefits.

The results lead us to the following practical suggestions regarding the groupings. For a one-year growth forecast, using all available information (i.e., the `grouped all' version) is beneficial as it is the best or close to the best performer across the different weights. For the same reason, we also recommend this grouping for two-year inflation and two-year unemployment forecasts. For one-year unemployment, one should group variables as this grouping is the best or close to the best across the different weights. For one-year inflation, grouped horizons deliver stable improvement across different weights (even though grouping variables works best for optimal weights). Finally, for a two-year growth forecast, we recommend grouping horizons but avoiding optimal weights. The convexity of the weights seems to be critical to avoid instabilities of negative weights, which was recently documented by \citet{Radchenko2023}.

If one is working with only a single forecasting horizon, the selection of grouping becomes redundant. Also, in other applications, grouping all tasks seems a sensible default provided $\gamma$ is chosen judiciously on a task-by-task basis. This default option can be improved by using additional cross-validation to help determine which grouping performs best.

\subsection{Tuned globalisation}
\label{sec:tuned}

The second set of experiments are broader comparisons that acknowledge the level of globalisation requires tuning in practice. For this purpose, we use leave-one-out cross-validation---a valid procedure provided the combination forecast errors are uncorrelated \citep{Bergmeir2018}. The value of $\gamma$ is tuned over ten values logarithmically spaced between $0.001$ and $1000$ on a per-task basis, so different tasks need not use the same value. To allow for comparisons of forecast accuracy across weighting schemes, we report the mean square forecast error relative to that from equal weights, a common benchmark in practice:
\begin{equation*}
\text{MSFE relative to equal}:=\frac{\sum_{t=\underaccent{\bar}{T}-h}^{\bar{T}-h}(y_{t+h}^{(k)}-\tilde{f}_{t+h\mid t}^{(k)})^2}{\sum_{t=\underaccent{\bar}{T}-h}^{\bar{T}-h}(y_{t+h}^{(k)}-\bar{f}_{t+h\mid t}^{(k)})^2},
\end{equation*}
where $\tilde{f}_{t+h\mid t}^{(k)}$ is an arbitrary combination forecast and $\bar{f}_{t+h\mid t}^{(k)}$ is the equally-weighted combination forecast. Values of this metric less than one indicate superior performance to equal weights.

Table~\ref{tab:spf-tuning} reports the average value of the performance metric across the six tasks, with the minimal and maximal values among the tasks in brackets. The shrinkage parameter $\lambda=0.1$. We study tuned $\lambda$ next. 
\begin{table}[ht!]
\centering
\footnotesize
\begin{tabularx}{390pt}{lRRR}
\toprule
 & Local combination & Hard global combination & Soft global combination \\ 
\midrule
\multicolumn{4}{c}{\emph{2017 Q1 to 2019 Q4}} \\
\multicolumn{4}{l}{\emph{Optimal weights}} \\ 
Grouped variables & 1.059 [0.272, 2.354] & 0.909 [0.512, 1.977] & 0.907 [0.359, 2.207] \\ 
Grouped horizons & 1.059 [0.272, 2.354] & 0.956 [0.306, 1.830] & 0.967 [0.299, 1.933] \\ 
Grouped all & 1.059 [0.272, 2.354] & 0.894 [0.592, 1.669] & 0.856 [0.360, 1.720] \\ 
\multicolumn{4}{l}{\emph{Optimal convex weights}} \\ 
Grouped variables & 0.991 [0.878, 1.159] & 1.033 [0.878, 1.410] & 0.968 [0.867, 1.111] \\ 
Grouped horizons & 0.991 [0.878, 1.159] & 0.957 [0.859, 1.178] & 0.981 [0.879, 1.177] \\ 
Grouped all & 0.991 [0.878, 1.159] & 1.001 [0.893, 1.237] & 0.980 [0.867, 1.183] \\ 
\multicolumn{4}{l}{\emph{Optimal equal weights}} \\ 
Grouped variables & 1.005 [0.861, 1.223] & 1.016 [0.861, 1.298] & 0.959 [0.848, 1.097] \\ 
Grouped horizons & 1.005 [0.861, 1.223] & 0.962 [0.873, 1.188] & 0.990 [0.866, 1.181] \\ 
Grouped all & 1.005 [0.861, 1.223] & 1.018 [0.880, 1.384] & 0.972 [0.850, 1.146] \\ 
\multicolumn{4}{c}{\emph{2020 Q1 to 2021 Q4}} \\
\multicolumn{4}{l}{\emph{Optimal weights}} \\ 
Grouped variables & 1.046 [0.913, 1.269] & 1.058 [0.959, 1.210] & 1.023 [0.931, 1.189] \\ 
Grouped horizons & 1.046 [0.913, 1.269] & 1.002 [0.784, 1.171] & 0.994 [0.718, 1.156] \\ 
Grouped all & 1.046 [0.913, 1.269] & 1.007 [0.959, 1.078] & 0.995 [0.858, 1.089] \\ 
\multicolumn{4}{l}{\emph{Optimal convex weights}} \\ 
Grouped variables & 1.006 [0.953, 1.072] & 0.953 [0.815, 1.032] & 0.992 [0.907, 1.044] \\ 
Grouped horizons & 1.006 [0.953, 1.072] & 1.009 [0.947, 1.048] & 1.005 [0.957, 1.051] \\ 
Grouped all & 1.006 [0.953, 1.072] & 0.931 [0.764, 1.009] & 0.999 [0.947, 1.042] \\ 
\multicolumn{4}{l}{\emph{Optimal equal weights}} \\ 
Grouped variables & 1.012 [0.942, 1.112] & 0.970 [0.772, 1.082] & 1.017 [0.949, 1.112] \\ 
Grouped horizons & 1.012 [0.942, 1.112] & 1.029 [0.959, 1.140] & 1.023 [0.954, 1.112] \\ 
Grouped all & 1.012 [0.942, 1.112] & 0.934 [0.681, 1.027] & 0.997 [0.920, 1.112] \\ 
\bottomrule
\end{tabularx}
\caption{Mean square forecast errors for the Survey of Professional Forecasters with cross-validated globalisation parameter. Averages over all tasks are next to minimums and maximums over all tasks in brackets. All values are relative to equal weights. The shrinkage parameter $\lambda=0.1$.}
\label{tab:spf-tuning}
\end{table}
The last five years of the data is again studied, but we now include the period 2020 Q1 to 2021 Q4 to evaluate recent performance during the COVID-19 recession and 2021--2022 inflation surge. Figure~\ref{fig:spf-data} highlights how the quarters on and after 2020 Q1 contain several outliers. To prevent these outliers dominating the performance metric, the testing set is split before and after 2020 Q1. Likewise, to avoid the outliers contaminating the estimated covariance matrices and thus the estimated weights, the training set is stopped at 2019 Q4.

With few exceptions, soft global combination improves on local combination. The improvements are generally greatest pre-2020. The more minor improvements post-2020 are possibly a consequence of the recent period of deteriorated economic conditions during which task-relatedness could be less stable. In some instances, hard global combination outperforms both soft global combination and local combination. However, as in the previous section, it also sometimes underperforms. On the other hand, the data-driven determination of the globalisation level for soft global combination produces good combinations that consistently forecast well.

Optimal weights realise the most significant gains from globalisation among the three weighting schemes---soft global combination (grouped all) places first in terms of average performance across tasks (pre-2020) compared with local combination, which places last. Moreover, globalisation leads to smaller maximal loss for optimal weights. Though not always beating optimal weights according to average performance, optimal convex weights and optimal equal weights have more consistent performance across tasks, especially pre-2020. With a suitable amount of globalisation, each weighting scheme can beat the notoriously difficult benchmark of equal weights for one or more task groupings.

\subsection{Tuned shrinkage}

The results of Table~\ref{tab:spf-tuning} are from tuning the globalisation parameter $\gamma$ while holding the shrinkage parameter $\lambda$ fixed. It is insightful to evaluate whether there are further benefits from tuning $\lambda$ in addition to $\gamma$. To this end, we cross-validate both parameters here. We focus on optimal (convex) weights to keep computation time reasonably low. Table~\ref{tab:spf-tuning-shrinkage} reports the results.
\begin{table}[ht!]
\centering
\footnotesize
\begin{tabularx}{390pt}{lRRR}
\toprule
 & Local combination & Hard global combination & Soft global combination \\ 
\midrule
\multicolumn{4}{c}{\emph{2017 Q1 to 2019 Q4}} \\
\multicolumn{4}{l}{\emph{Optimal weights}} \\ 
Grouped variables & 0.843 [0.418, 1.493] & 0.859 [0.495, 1.494] & 0.774 [0.418, 1.535] \\ 
Grouped horizons & 0.843 [0.418, 1.493] & 0.832 [0.474, 1.237] & 0.825 [0.421, 1.418] \\ 
Grouped all & 0.843 [0.418, 1.493] & 0.877 [0.547, 1.523] & 0.800 [0.418, 1.562] \\ 
\multicolumn{4}{l}{\emph{Optimal convex weights}} \\ 
Grouped variables & 1.004 [0.908, 1.104] & 1.016 [0.865, 1.224] & 0.982 [0.883, 1.120] \\ 
Grouped horizons & 1.004 [0.908, 1.104] & 0.978 [0.908, 1.134] & 0.979 [0.906, 1.104] \\ 
Grouped all & 1.004 [0.908, 1.104] & 0.999 [0.885, 1.233] & 0.989 [0.902, 1.156] \\ 
\multicolumn{4}{c}{\emph{2020 Q1 to 2021 Q4}} \\
\multicolumn{4}{l}{\emph{Optimal weights}} \\ 
Grouped variables & 0.968 [0.833, 1.025] & 1.030 [0.972, 1.106] & 0.980 [0.832, 1.074] \\ 
Grouped horizons & 0.968 [0.833, 1.025] & 0.936 [0.751, 1.016] & 0.968 [0.843, 1.019] \\ 
Grouped all & 0.968 [0.833, 1.025] & 0.997 [0.942, 1.053] & 0.973 [0.810, 1.063] \\ 
\multicolumn{4}{l}{\emph{Optimal convex weights}} \\ 
Grouped variables & 1.001 [0.969, 1.072] & 0.921 [0.715, 1.025] & 0.968 [0.825, 1.034] \\ 
Grouped horizons & 1.001 [0.969, 1.072] & 0.995 [0.911, 1.093] & 1.004 [0.969, 1.086] \\ 
Grouped all & 1.001 [0.969, 1.072] & 0.964 [0.870, 1.032] & 0.994 [0.969, 1.025] \\ 
\bottomrule
\end{tabularx}

\caption{Mean square forecast errors for the Survey of Professional Forecasters with cross-validated globalisation and shrinkage parameters. Averages over all tasks are next to minimums and maximums over all tasks in brackets. All values are relative to equal weights.}
\label{tab:spf-tuning-shrinkage}
\end{table}
Optimal weights witness an improvement across the board relative to the results of Table~\ref{tab:spf-tuning} (for local, hard global, and soft global combinations). Though it is known \citep[see][]{Roccazzella2022} that optimal weights benefit from (carefully tuned) shrinkage, our result is the first documentation of similar behaviour for global combination. The results for optimal convex weights---whose nonnegativity constraint already imparts a form of shrinkage---are similar to Table~\ref{tab:spf-tuning}. Our core finding remains the same in both cases: globalisation via soft global combination is typically beneficial.

\subsection{Forecast combination puzzle}

In more than 50 years of forecast combination literature spanning a myriad of weighting schemes, `forecasters still have little guidance on how to solve the forecast combination puzzle' \citep{Wang2023}. Our empirical results show that before the COVID-19 recession (2017--2019), soft global combination not only improves upon local and hard global combinations but also upon equal weights. Our synthetic experiments also demonstrate improvement in stable conditions; see \ref{app:equal}. However, results post-2019 are mixed with around half cases where soft global combination performs better than equal weights.

One possible explanation is a structural break produced by the COVID-19 recession. \citet{Rossi2001} showed that the 2007--2008 global financial crisis (GFC) affected forecasting performance significantly. Currently, no similar research is available for the COVID period. However, Figure~\ref{fig:spf-data} leaves no doubt that for inflation and growth, the effects of COVID far exceed those observed during the GFC. Simple tests for a structural break in Table~\ref{tab:spf-structural-break} support this claim.
\begin{table}[ht!]
\centering
\footnotesize
\begin{tabular}{lrr}
\toprule
 & \multicolumn{2}{c}{$p$-value} \\ 
\cmidrule(lr){2-3}
 & $t$ test & $F$ test \\ 
\midrule
One-year growth & 0.222 & <0.001 \\ 
Two-year growth & 0.631 & <0.001 \\ 
One-year inflation & 0.354 & <0.001 \\ 
Two-year inflation & 0.559 & <0.001 \\ 
One-year unemployment & 0.127 & 0.068 \\ 
Two-year unemployment & 0.556 & 0.132 \\ 
\bottomrule
\end{tabular}

\caption{Tests for a structural break in the average forecast error between pre- and post-COVID periods (up to 2019 Q4 and after 2019 Q4). $p$-values are reported from a $t$ test for a difference in means and an $F$ test for a difference in variances.}
\label{tab:spf-structural-break}
\end{table}
The variance of the average forecast error is significantly different post-COVID at the 1\% level (except for the unemployment forecasts). As there is little data available post-COVID, reestimated weights are likely to have large variability that negates the benefits of soft global combination. If weights from the pre-COVID period are used, they will not necessarily be optimal and may not provide the benefits observed under stable conditions. \citet{Wang2023} recommend equal weights in such cases. Until more data is available, equal weights are probably suitable for the post-COVID period. With more post-COVID data, soft global combination should quickly catch up as a strong competitor and a potential solution to the forecast combination puzzle; see also \citet{Frazier2023}.

The benefits of equal weights centre around the substantial reduction of the variance at the cost of introducing a small bias; see \citet{Claeskens2016}. A more recent approach by \citet{Blanc2020} requires an explicit solution to analyse the bias-variance trade-off. With the absence of an explicit solution in our case, a practitioner needs to empirically validate whether soft global combination beats the equally-weighted combination weights in their setting. Our findings suggest the likelihood of improving over the equal weights is high.

\section{Concluding remarks}
\label{sec:conclusion}

To date, the problem of combining economic forecasts has been handled on a per-task basis, with the combination for each variable and forecast horizon learned independently of other variables and horizons. When the forecasting tasks are related, as economic theory and evidence suggest, this approach of learning the combinations using only local information is potentially suboptimal. This paper investigates the value of a global approach, where task-relatedness is directly exploited to improve the quality of combinations. At the heart of our approach is a principled framework that accounts for the level of homogeneity across tasks by flexibly interpolating between fully local and fully global combinations. In addition to unifying local and global approaches under one umbrella, the new framework accommodates many existing weighting schemes. Empirical evidence from the European Central Bank SPF suggests combinations of expert forecasts for rates of growth, inflation, and unemployment in the Eurozone benefit from some degree of globalisation, as do combinations of these same variables across one- and two-year horizons. Further empirical evidence on economic and financial data from the M4 Competition in \ref{app:m4} indicates similar benefits for combinations of model-based forecasts.

Our approach is not limited to point forecasts and can be extended to probabilistic forecasts. Consider, e.g., the optimal weights of \citet{Hall2007} and \citet{Geweke2011} that, in the case of only one task, maximise the log-score when combining $p$ individual predictive densities $p_{j} (\cdot; \bm{\theta}_j)$ using the historical observations $y_1,\dots,y_T$:
\begin{equation*}
\underset{\bm{w}\in\mathcal{W}^\mathrm{optcvx}}{\max}\,\sum_{t=1}^T\log \left[ \sum_{j=1}^pw_jp_j(y_t;\bm{\theta}_j)\right].
\end{equation*}
These weights readily extend to soft global combination for $m$ tasks:
\begin{equation*}
\underset{\substack{\bm{w}^{(1)},\ldots,\bm{w}^{(m)}\in\mathcal{W}^\mathrm{optcvx} \\ \bar{\bm{w}}\in\mathbb{R}^p}}{\max}\,\sum_{k=1}^m\left(\sum_{t=1}^T\log\left[ \sum_{j=1}^pw_j^{(k)} p_j(y_t^{(k)};\bm{\theta}_j^{(k)})\right]+\gamma\|\bar{\bm{w}}-\bm{w}^{(k)}\|_q^q\right),\quad q\in\{1,2\}.
\end{equation*}
The density parameter $\bm{\theta}_{j}^{(k)}$ can be different across tasks. This problem can be further enhanced using a shrinkage penalty or additional constraints, e.g., the high moment constraints of \citet{Pauwels2023}.

Furthermore, our approach is based on the intuitive idea that a forecaster’s competence in predicting one variable might contain some signal about their competence in predicting another. One can examine the connection between the accuracy of the combined forecast and individual forecaster characteristics. We leave this direction for future research.

An \texttt{R} implementation of the global forecast combinations in this paper is publicly available at
\begin{center}
\url{https://github.com/ryan-thompson/global-combinations}.
\end{center}

\section*{Acknowledgements}

We are especially grateful to the editor, associate editor, and two referees for their thoughtful, constructive, and encouraging reviews that greatly improved the manuscript. We are also grateful for valuable comments from the participants of the 43rd International Symposium on Forecasting, 6th International Conference on Econometrics and Statistics, and 2023 Econometric Society Australasian Meeting. Ryan Thompson acknowledges financial support by an Australian Government Research Training Program (RTP) Scholarship through Monash University, where he worked on the first draft of this paper as a PhD student.

\section*{Conflict of interest}

The authors declare no conflict of interest.

\bibliographystyle{elsarticle-harv}
\bibliography{library}

\begin{thebibliography}{55}
\expandafter\ifx\csname natexlab\endcsname\relax\def\natexlab#1{#1}\fi
\providecommand{\url}[1]{\texttt{#1}}
\providecommand{\href}[2]{#2}
\providecommand{\path}[1]{#1}
\providecommand{\DOIprefix}{doi:}
\providecommand{\ArXivprefix}{arXiv:}
\providecommand{\URLprefix}{URL: }
\providecommand{\Pubmedprefix}{pmid:}
\providecommand{\doi}[1]{\href{http://dx.doi.org/#1}{\path{#1}}}
\providecommand{\Pubmed}[1]{\href{pmid:#1}{\path{#1}}}
\providecommand{\bibinfo}[2]{#2}
\ifx\xfnm\relax \def\xfnm[#1]{\unskip,\space#1}\fi
\bibitem[{Aiolfi and Timmermann(2006)}]{Aiolfi2006}
\bibinfo{author}{Aiolfi, M.}, \bibinfo{author}{Timmermann, A.},
  \bibinfo{year}{2006}.
\newblock \bibinfo{title}{Persistence in forecasting performance and
  conditional combination strategies}.
\newblock \bibinfo{journal}{Journal of Econometrics} \bibinfo{volume}{135},
  \bibinfo{pages}{31--53}.
\bibitem[{Bates et~al.(2022)Bates, Maechler and Jagan}]{Bates2022}
\bibinfo{author}{Bates, D.}, \bibinfo{author}{Maechler, M.},
  \bibinfo{author}{Jagan, M.}, \bibinfo{year}{2022}.
\newblock \bibinfo{title}{Matrix: Sparse and dense matrix classes and methods}.
\newblock \URLprefix \url{https://CRAN.R-project.org/package=Matrix}.
  \bibinfo{note}{{R} package version 1.5-3}.
\bibitem[{Bates and Granger(1969)}]{Bates1969}
\bibinfo{author}{Bates, J.M.}, \bibinfo{author}{Granger, C.W.J.},
  \bibinfo{year}{1969}.
\newblock \bibinfo{title}{The combination of forecasts}.
\newblock \bibinfo{journal}{OR} \bibinfo{volume}{20},
  \bibinfo{pages}{451--468}.
\bibitem[{Bergmeir et~al.(2018)Bergmeir, Hyndman and Koo}]{Bergmeir2018}
\bibinfo{author}{Bergmeir, C.}, \bibinfo{author}{Hyndman, R.J.},
  \bibinfo{author}{Koo, B.}, \bibinfo{year}{2018}.
\newblock \bibinfo{title}{A note on the validity of cross-validation for
  evaluating autoregressive time series prediction}.
\newblock \bibinfo{journal}{Computational Statistics and Data Analysis}
  \bibinfo{volume}{120}, \bibinfo{pages}{70--83}.
\bibitem[{Blanc and Setzer(2020)}]{Blanc2020}
\bibinfo{author}{Blanc, S.M.}, \bibinfo{author}{Setzer, T.},
  \bibinfo{year}{2020}.
\newblock \bibinfo{title}{Bias-variance trade-off and shrinkage of weights in
  forecast combination}.
\newblock \bibinfo{journal}{Management Science} \bibinfo{volume}{66},
  \bibinfo{pages}{5720--5737}.
\bibitem[{B{\"{u}}rgi and Sinclair(2017)}]{Burgi2017}
\bibinfo{author}{B{\"{u}}rgi, C.}, \bibinfo{author}{Sinclair, T.M.},
  \bibinfo{year}{2017}.
\newblock \bibinfo{title}{A nonparametric approach to identifying a subset of
  forecasters that outperforms the simple average}.
\newblock \bibinfo{journal}{Empirical Economics} \bibinfo{volume}{53},
  \bibinfo{pages}{101--115}.
\bibitem[{Capistr{\'{a}}n and Timmermann(2009)}]{Capistran2009}
\bibinfo{author}{Capistr{\'{a}}n, C.}, \bibinfo{author}{Timmermann, A.},
  \bibinfo{year}{2009}.
\newblock \bibinfo{title}{Forecast combination with entry and exit of experts}.
\newblock \bibinfo{journal}{Journal of Business and Economic Statistics}
  \bibinfo{volume}{27}, \bibinfo{pages}{428--440}.
\bibitem[{Caruana(1997)}]{Caruana1997}
\bibinfo{author}{Caruana, R.}, \bibinfo{year}{1997}.
\newblock \bibinfo{title}{Multitask learning}.
\newblock \bibinfo{journal}{Machine Learning} \bibinfo{volume}{28},
  \bibinfo{pages}{41--75}.
\bibitem[{Claeskens et~al.(2016)Claeskens, Magnus, Vasnev and
  Wang}]{Claeskens2016}
\bibinfo{author}{Claeskens, G.}, \bibinfo{author}{Magnus, J.R.},
  \bibinfo{author}{Vasnev, A.L.}, \bibinfo{author}{Wang, W.},
  \bibinfo{year}{2016}.
\newblock \bibinfo{title}{The forecast combination puzzle: A simple theoretical
  explanation}.
\newblock \bibinfo{journal}{International Journal of Forecasting}
  \bibinfo{volume}{32}, \bibinfo{pages}{754--762}.
\bibitem[{Conflitti et~al.(2015)Conflitti, {De Mol} and
  Giannone}]{Conflitti2015}
\bibinfo{author}{Conflitti, C.}, \bibinfo{author}{{De Mol}, C.},
  \bibinfo{author}{Giannone, D.}, \bibinfo{year}{2015}.
\newblock \bibinfo{title}{Optimal combination of survey forecasts}.
\newblock \bibinfo{journal}{International Journal of Forecasting}
  \bibinfo{volume}{31}, \bibinfo{pages}{1096--1103}.
\bibitem[{Diebold and Shin(2019)}]{Diebold2019}
\bibinfo{author}{Diebold, F.X.}, \bibinfo{author}{Shin, M.},
  \bibinfo{year}{2019}.
\newblock \bibinfo{title}{Machine learning for regularized survey forecast
  combination: Partially-egalitarian lasso and its derivatives}.
\newblock \bibinfo{journal}{International Journal of Forecasting}
  \bibinfo{volume}{35}, \bibinfo{pages}{1679--1691}.
\bibitem[{Evgeniou et~al.(2005)Evgeniou, Micchelli and Pontil}]{Evgeniou2005}
\bibinfo{author}{Evgeniou, T.}, \bibinfo{author}{Micchelli, C.A.},
  \bibinfo{author}{Pontil, M.}, \bibinfo{year}{2005}.
\newblock \bibinfo{title}{Learning multiple tasks with kernel methods}.
\newblock \bibinfo{journal}{Journal of Machine Learning Research}
  \bibinfo{volume}{6}, \bibinfo{pages}{615--637}.
\bibitem[{Evgeniou and Pontil(2004)}]{Evgeniou2004}
\bibinfo{author}{Evgeniou, T.}, \bibinfo{author}{Pontil, M.},
  \bibinfo{year}{2004}.
\newblock \bibinfo{title}{Regularized multi--task learning}, in:
  \bibinfo{booktitle}{Proceedings of the 10th ACM SIGKDD International
  Conference on Knowledge Discovery and Data Mining}, pp.
  \bibinfo{pages}{109--117}.
\bibitem[{Frazier et~al.(2023)Frazier, Covey, Martin and Poskitt}]{Frazier2023}
\bibinfo{author}{Frazier, D.T.}, \bibinfo{author}{Covey, R.},
  \bibinfo{author}{Martin, G.M.}, \bibinfo{author}{Poskitt, D.},
  \bibinfo{year}{2023}.
\newblock \bibinfo{title}{Solving the forecast combination puzzle}
  \href{http://arxiv.org/abs/2308.05263}{{\tt arXiv:2308.05263}}.
\bibitem[{Genre et~al.(2013)Genre, Kenny, Meyler and Timmermann}]{Genre2013}
\bibinfo{author}{Genre, V.}, \bibinfo{author}{Kenny, G.},
  \bibinfo{author}{Meyler, A.}, \bibinfo{author}{Timmermann, A.},
  \bibinfo{year}{2013}.
\newblock \bibinfo{title}{Combining expert forecasts: Can anything beat the
  simple average?}
\newblock \bibinfo{journal}{International Journal of Forecasting}
  \bibinfo{volume}{29}, \bibinfo{pages}{108--121}.
\bibitem[{Geweke and Amisano(2011)}]{Geweke2011}
\bibinfo{author}{Geweke, J.}, \bibinfo{author}{Amisano, G.},
  \bibinfo{year}{2011}.
\newblock \bibinfo{title}{Optimal prediction pools}.
\newblock \bibinfo{journal}{Journal of Econometrics} \bibinfo{volume}{164},
  \bibinfo{pages}{130--141}.
\bibitem[{Gibbs and Vasnev(2021)}]{Gibbs2021}
\bibinfo{author}{Gibbs, C.G.}, \bibinfo{author}{Vasnev, A.L.},
  \bibinfo{year}{2021}.
\newblock \bibinfo{title}{Conditionally optimal weights and forward-looking
  approaches to combining forecasts} \URLprefix
  \url{https://christopherggibbs.weebly.com/uploads/3/8/2/6/38260553/cow_2021_forweb.pdf}.
\bibitem[{Godahewa et~al.(2021)Godahewa, Bandara, Webb, Smyl and
  Bergmeir}]{Godahewa2021}
\bibinfo{author}{Godahewa, R.}, \bibinfo{author}{Bandara, K.},
  \bibinfo{author}{Webb, G.I.}, \bibinfo{author}{Smyl, S.},
  \bibinfo{author}{Bergmeir, C.}, \bibinfo{year}{2021}.
\newblock \bibinfo{title}{Ensembles of localised models for time series
  forecasting}.
\newblock \bibinfo{journal}{Knowledge-Based Systems} \bibinfo{volume}{233},
  \bibinfo{pages}{107518}.
\bibitem[{Granger and Ramanathan(1984)}]{Granger1984}
\bibinfo{author}{Granger, C.W.J.}, \bibinfo{author}{Ramanathan, R.},
  \bibinfo{year}{1984}.
\newblock \bibinfo{title}{Improved methods of combining forecasts}.
\newblock \bibinfo{journal}{Journal of Forecasting} \bibinfo{volume}{3},
  \bibinfo{pages}{197--204}.
\bibitem[{{Gurobi Optimization, LLC}(2023)}]{gurobi2023}
\bibinfo{author}{{Gurobi Optimization, LLC}}, \bibinfo{year}{2023}.
\newblock \bibinfo{title}{{Gurobi Optimizer} reference manual}.
\newblock \URLprefix \url{https://www.gurobi.com}.
\bibitem[{Hall and Mitchell(2007)}]{Hall2007}
\bibinfo{author}{Hall, S.G.}, \bibinfo{author}{Mitchell, J.},
  \bibinfo{year}{2007}.
\newblock \bibinfo{title}{Combining density forecasts}.
\newblock \bibinfo{journal}{International Journal of Forecasting}
  \bibinfo{volume}{23}, \bibinfo{pages}{1--13}.
\bibitem[{Hansen(2008)}]{Hansen2008}
\bibinfo{author}{Hansen, B.E.}, \bibinfo{year}{2008}.
\newblock \bibinfo{title}{Least-squares forecast averaging}.
\newblock \bibinfo{journal}{Journal of Econometrics} \bibinfo{volume}{146},
  \bibinfo{pages}{342--350}.
\bibitem[{Hoerl and Kennard(1970)}]{Hoerl1970}
\bibinfo{author}{Hoerl, A.E.}, \bibinfo{author}{Kennard, R.W.},
  \bibinfo{year}{1970}.
\newblock \bibinfo{title}{Ridge regression: Biased estimation for nonorthogonal
  problems}.
\newblock \bibinfo{journal}{Technometrics} \bibinfo{volume}{12},
  \bibinfo{pages}{55--67}.
\bibitem[{Hyndman et~al.(2023)Hyndman, Athanasopoulos, Bergmeir, Caceres,
  Chhay, O'Hara-Wild, Petropoulos, Razbash, Wang and Yasmeen}]{Hyndman2023}
\bibinfo{author}{Hyndman, R.}, \bibinfo{author}{Athanasopoulos, G.},
  \bibinfo{author}{Bergmeir, C.}, \bibinfo{author}{Caceres, G.},
  \bibinfo{author}{Chhay, L.}, \bibinfo{author}{O'Hara-Wild, M.},
  \bibinfo{author}{Petropoulos, F.}, \bibinfo{author}{Razbash, S.},
  \bibinfo{author}{Wang, E.}, \bibinfo{author}{Yasmeen, F.},
  \bibinfo{year}{2023}.
\newblock \bibinfo{title}{forecast: Forecasting functions for time series and
  linear models}.
\newblock \URLprefix \url{https://pkg.robjhyndman.com/forecast/}.
  \bibinfo{note}{{R} package version 8.21}.
\bibitem[{Kourentzes et~al.(2019)Kourentzes, Barrow and
  Petropoulos}]{Kourentzes2019}
\bibinfo{author}{Kourentzes, N.}, \bibinfo{author}{Barrow, D.},
  \bibinfo{author}{Petropoulos, F.}, \bibinfo{year}{2019}.
\newblock \bibinfo{title}{Another look at forecast selection and combination:
  Evidence from forecast pooling}.
\newblock \bibinfo{journal}{International Journal of Production Economics}
  \bibinfo{volume}{209}, \bibinfo{pages}{226--235}.
\bibitem[{Lahti et~al.(2017)Lahti, Huovari, Kainu and Biecek}]{Lahti2017}
\bibinfo{author}{Lahti, L.}, \bibinfo{author}{Huovari, J.},
  \bibinfo{author}{Kainu, M.}, \bibinfo{author}{Biecek, P.},
  \bibinfo{year}{2017}.
\newblock \bibinfo{title}{Retrieval and analysis of {Eurostat} open data with
  the eurostat package}.
\newblock \bibinfo{journal}{R Journal} \bibinfo{volume}{9},
  \bibinfo{pages}{385--392}.
\bibitem[{Laptev et~al.(2017)Laptev, Yosinski, Li and Smyl}]{Laptev2017}
\bibinfo{author}{Laptev, N.}, \bibinfo{author}{Yosinski, J.},
  \bibinfo{author}{Li, L.E.}, \bibinfo{author}{Smyl, S.}, \bibinfo{year}{2017}.
\newblock \bibinfo{title}{Time-series extreme event forecasting with neural
  networks at {Uber}}, in: \bibinfo{booktitle}{ICML 2017 Time Series Workshop}.
\bibitem[{Lawrence et~al.(2006)Lawrence, Goodwin, O'Connor and
  \"{O}nkal}]{Lawrence2006}
\bibinfo{author}{Lawrence, M.}, \bibinfo{author}{Goodwin, P.},
  \bibinfo{author}{O'Connor, M.}, \bibinfo{author}{\"{O}nkal, D.},
  \bibinfo{year}{2006}.
\newblock \bibinfo{title}{Judgmental forecasting: A review of progress over the
  last 25 years}.
\newblock \bibinfo{journal}{International Journal of Forecasting}
  \bibinfo{volume}{22}, \bibinfo{pages}{493--518}.
\bibitem[{Ledoit and Wolf(2004)}]{Ledoit2004}
\bibinfo{author}{Ledoit, O.}, \bibinfo{author}{Wolf, M.}, \bibinfo{year}{2004}.
\newblock \bibinfo{title}{A well-conditioned estimator for large-dimensional
  covariance matrices}.
\newblock \bibinfo{journal}{Journal of Multivariate Analysis}
  \bibinfo{volume}{88}, \bibinfo{pages}{365--411}.
\bibitem[{Liberti(2008)}]{Liberti2008}
\bibinfo{author}{Liberti, L.}, \bibinfo{year}{2008}.
\newblock \bibinfo{title}{{Introduction to global optimization}}.
\newblock \bibinfo{type}{Technical Report}.
\newblock \URLprefix
  \url{https://www.lix.polytechnique.fr/~liberti/teaching/globalopt-lima.pdf}.
\bibitem[{Magnus and Vasnev(2023)}]{Magnus2023}
\bibinfo{author}{Magnus, J.R.}, \bibinfo{author}{Vasnev, A.L.},
  \bibinfo{year}{2023}.
\newblock \bibinfo{title}{On the uncertainty of a combined forecast: The
  critical role of correlation}.
\newblock \bibinfo{journal}{International Journal of Forecasting}
  \bibinfo{volume}{39}, \bibinfo{pages}{1895--1908}.
\bibitem[{Makridakis et~al.(2020)Makridakis, Spiliotis and
  Assimakopoulos}]{Makridakis2020}
\bibinfo{author}{Makridakis, S.}, \bibinfo{author}{Spiliotis, E.},
  \bibinfo{author}{Assimakopoulos, V.}, \bibinfo{year}{2020}.
\newblock \bibinfo{title}{The {M4 Competition}: 100,000 time series and 61
  forecasting methods}.
\newblock \bibinfo{journal}{International Journal of Forecasting}
  \bibinfo{volume}{36}, \bibinfo{pages}{54--74}.
\bibitem[{Matsypura et~al.(2018)Matsypura, Thompson and Vasnev}]{Matsypura2018}
\bibinfo{author}{Matsypura, D.}, \bibinfo{author}{Thompson, R.},
  \bibinfo{author}{Vasnev, A.L.}, \bibinfo{year}{2018}.
\newblock \bibinfo{title}{Optimal selection of expert forecasts with integer
  programming}.
\newblock \bibinfo{journal}{Omega} \bibinfo{volume}{78},
  \bibinfo{pages}{165--175}.
\bibitem[{Montero-Manso et~al.(2020)Montero-Manso, Athanasopoulos, Hyndman and
  Talagala}]{Montero-Manso2020}
\bibinfo{author}{Montero-Manso, P.}, \bibinfo{author}{Athanasopoulos, G.},
  \bibinfo{author}{Hyndman, R.J.}, \bibinfo{author}{Talagala, T.S.},
  \bibinfo{year}{2020}.
\newblock \bibinfo{title}{{FFORMA}: Feature-based forecast model averaging}.
\newblock \bibinfo{journal}{International Journal of Forecasting}
  \bibinfo{volume}{36}, \bibinfo{pages}{86--92}.
\bibitem[{Montero-Manso and Hyndman(2021)}]{Montero-Manso2021}
\bibinfo{author}{Montero-Manso, P.}, \bibinfo{author}{Hyndman, R.J.},
  \bibinfo{year}{2021}.
\newblock \bibinfo{title}{Principles and algorithms for forecasting groups of
  time series: Locality and globality}.
\newblock \bibinfo{journal}{International Journal of Forecasting}
  \bibinfo{volume}{37}, \bibinfo{pages}{1632--1653}.
\bibitem[{Newbold and Granger(1974)}]{Newbold1974}
\bibinfo{author}{Newbold, P.}, \bibinfo{author}{Granger, C.W.J.},
  \bibinfo{year}{1974}.
\newblock \bibinfo{title}{Experience with forecasting univariate time series
  and the combination of forecasts}.
\newblock \bibinfo{journal}{Journal of the Royal Statistical Society: Series A
  (General)} \bibinfo{volume}{137}, \bibinfo{pages}{131--165}.
\bibitem[{Okun(1962)}]{Okun1962}
\bibinfo{author}{Okun, A.M.}, \bibinfo{year}{1962}.
\newblock \bibinfo{title}{Potential {GNP}: Its measurement and significance},
  in: \bibinfo{booktitle}{Proceedings of the Business and Economic Statistics
  Section of the American Statistical Association}.
\bibitem[{Pauwels et~al.(2023)Pauwels, Radchenko and Vasnev}]{Pauwels2023}
\bibinfo{author}{Pauwels, L.}, \bibinfo{author}{Radchenko, P.},
  \bibinfo{author}{Vasnev, A.L.}, \bibinfo{year}{2023}.
\newblock \bibinfo{title}{High moment constraints for predictive density
  combinations} \URLprefix
  \url{https://papers.ssrn.com/sol3/papers.cfm?abstract_id=3593124}.
\bibitem[{Persson(2022)}]{Persson2022}
\bibinfo{author}{Persson, E.}, \bibinfo{year}{2022}.
\newblock \bibinfo{title}{ecb: Programmatic access to the {European Central
  Bank's Statistical Data Warehouse}}.
\newblock \URLprefix \url{https://CRAN.R-project.org/package=ecb}.
  \bibinfo{note}{{R} package version 0.4.1}.
\bibitem[{Petropoulos et~al.(2018)Petropoulos, Kourentzes, Nikolopoulos and
  Siemsen}]{Petropoulos2018}
\bibinfo{author}{Petropoulos, F.}, \bibinfo{author}{Kourentzes, N.},
  \bibinfo{author}{Nikolopoulos, K.}, \bibinfo{author}{Siemsen, E.},
  \bibinfo{year}{2018}.
\newblock \bibinfo{title}{Judgmental selection of forecasting models}.
\newblock \bibinfo{journal}{Journal of Operations Management}
  \bibinfo{volume}{60}, \bibinfo{pages}{34--46}.
\bibitem[{Phillips(1958)}]{Phillips1958}
\bibinfo{author}{Phillips, A.W.}, \bibinfo{year}{1958}.
\newblock \bibinfo{title}{The relation between unemployment and the rate of
  change of money wage rates in the united kingdom, 1861–1957}.
\newblock \bibinfo{journal}{Economica} \bibinfo{volume}{25},
  \bibinfo{pages}{283--299}.
\bibitem[{Poncela et~al.(2011)Poncela, Rodr{\'{i}}guez, S{\'{a}}nchez-Mangas
  and Senra}]{Poncela2011}
\bibinfo{author}{Poncela, P.}, \bibinfo{author}{Rodr{\'{i}}guez, J.},
  \bibinfo{author}{S{\'{a}}nchez-Mangas, R.}, \bibinfo{author}{Senra, E.},
  \bibinfo{year}{2011}.
\newblock \bibinfo{title}{Forecast combination through dimension reduction
  techniques}.
\newblock \bibinfo{journal}{International Journal of Forecasting}
  \bibinfo{volume}{27}, \bibinfo{pages}{224--237}.
\bibitem[{Qian et~al.(2022)Qian, Rolling, Cheng and Yang}]{Qian2022}
\bibinfo{author}{Qian, W.}, \bibinfo{author}{Rolling, C.A.},
  \bibinfo{author}{Cheng, G.}, \bibinfo{author}{Yang, Y.},
  \bibinfo{year}{2022}.
\newblock \bibinfo{title}{Combining forecasts for universally optimal
  performance}.
\newblock \bibinfo{journal}{International Journal of Forecasting}
  \bibinfo{volume}{38}, \bibinfo{pages}{193--208}.
\bibitem[{{R Core Team}(2023)}]{R2023}
\bibinfo{author}{{R Core Team}}, \bibinfo{year}{2023}.
\newblock \bibinfo{title}{R: A language and environment for statistical
  computing}.
\newblock \bibinfo{organization}{R Foundation for Statistical Computing}.
  \bibinfo{address}{Vienna, Austria}.
\newblock \URLprefix \url{https://www.R-project.org/}.
\bibitem[{Radchenko et~al.(2023)Radchenko, Vasnev and Wang}]{Radchenko2023}
\bibinfo{author}{Radchenko, P.}, \bibinfo{author}{Vasnev, A.L.},
  \bibinfo{author}{Wang, W.}, \bibinfo{year}{2023}.
\newblock \bibinfo{title}{Too similar to combine? on negative weights in
  forecast combination}.
\newblock \bibinfo{journal}{International Journal of Forecasting}
  \bibinfo{volume}{39}, \bibinfo{pages}{18--38}.
\bibitem[{Roccazzella et~al.(2022)Roccazzella, Gambetti and
  Vrins}]{Roccazzella2022}
\bibinfo{author}{Roccazzella, F.}, \bibinfo{author}{Gambetti, P.},
  \bibinfo{author}{Vrins, F.}, \bibinfo{year}{2022}.
\newblock \bibinfo{title}{Optimal and robust combination of forecasts via
  constrained optimization and shrinkage}.
\newblock \bibinfo{journal}{International Journal of Forecasting}
  \bibinfo{volume}{38}, \bibinfo{pages}{97--116}.
\bibitem[{Rossi(2021)}]{Rossi2001}
\bibinfo{author}{Rossi, B.}, \bibinfo{year}{2021}.
\newblock \bibinfo{title}{Forecasting in the presence of instabilities: How we
  know whether models predict well and how to improve them}.
\newblock \bibinfo{journal}{Journal of Economic Literature}
  \bibinfo{volume}{59}, \bibinfo{pages}{1135--1190}.
\bibitem[{Salinas et~al.(2020)Salinas, Flunkert, Gasthaus and
  Januschowski}]{Salinas2020}
\bibinfo{author}{Salinas, D.}, \bibinfo{author}{Flunkert, V.},
  \bibinfo{author}{Gasthaus, J.}, \bibinfo{author}{Januschowski, T.},
  \bibinfo{year}{2020}.
\newblock \bibinfo{title}{{DeepAR}: Probabilistic forecasting with
  autoregressive recurrent networks}.
\newblock \bibinfo{journal}{International Journal of Forecasting}
  \bibinfo{volume}{36}, \bibinfo{pages}{1181--1191}.
\bibitem[{Sch{\"{a}}fer and Strimmer(2005)}]{Schafer2005}
\bibinfo{author}{Sch{\"{a}}fer, J.}, \bibinfo{author}{Strimmer, K.},
  \bibinfo{year}{2005}.
\newblock \bibinfo{title}{A shrinkage approach to large-scale covariance matrix
  estimation and implications for functional genomics}.
\newblock \bibinfo{journal}{Statistical Applications in Genetics and Molecular
  Biology} \bibinfo{volume}{4}.
\bibitem[{Stock and Watson(2004)}]{Stock2004}
\bibinfo{author}{Stock, J.H.}, \bibinfo{author}{Watson, M.W.},
  \bibinfo{year}{2004}.
\newblock \bibinfo{title}{Combination forecasts of output growth in a
  seven-country data set}.
\newblock \bibinfo{journal}{Journal of Forecasting} \bibinfo{volume}{23},
  \bibinfo{pages}{405--430}.
\bibitem[{Tibshirani(1996)}]{Tibshirani1996}
\bibinfo{author}{Tibshirani, R.}, \bibinfo{year}{1996}.
\newblock \bibinfo{title}{Regression shrinkage and selection via the lasso}.
\newblock \bibinfo{journal}{Journal of the Royal Statistical Society: Series B
  (Methodological)} \bibinfo{volume}{58}, \bibinfo{pages}{267--288}.
\bibitem[{Touloumis(2015)}]{Touloumis2015}
\bibinfo{author}{Touloumis, A.}, \bibinfo{year}{2015}.
\newblock \bibinfo{title}{Nonparametric {Stein}-type shrinkage covariance
  matrix estimators in high-dimensional settings}.
\newblock \bibinfo{journal}{Computational Statistics and Data Analysis}
  \bibinfo{volume}{83}, \bibinfo{pages}{251--261}.
\bibitem[{Wang et~al.(2023)Wang, Hyndman, Li and Kang}]{Wang2023}
\bibinfo{author}{Wang, X.}, \bibinfo{author}{Hyndman, R.J.},
  \bibinfo{author}{Li, F.}, \bibinfo{author}{Kang, Y.}, \bibinfo{year}{2023}.
\newblock \bibinfo{title}{Forecast combinations: An over 50-year review}.
\newblock \bibinfo{journal}{International Journal of Forecasting}
  \bibinfo{volume}{39}, \bibinfo{pages}{1518--1547}.
\bibitem[{Yang(2004)}]{Yang2004}
\bibinfo{author}{Yang, Y.}, \bibinfo{year}{2004}.
\newblock \bibinfo{title}{Combining forecasting procedures: Some theoretical
  results}.
\newblock \bibinfo{journal}{Econometric Theory} \bibinfo{volume}{20},
  \bibinfo{pages}{176--222}.
\bibitem[{Zhang and Yang(2022)}]{Zhang2022}
\bibinfo{author}{Zhang, Y.}, \bibinfo{author}{Yang, Q.}, \bibinfo{year}{2022}.
\newblock \bibinfo{title}{A survey on multi-task learning}.
\newblock \bibinfo{journal}{IEEE Transactions on Knowledge and Data
  Engineering} \bibinfo{volume}{34}, \bibinfo{pages}{5586--5609}.

\end{thebibliography}

\appendix

\section{Proofs}
\label{app:proofs}

\subsection{Proof of Proposition~\ref{prop:extra}}
\label{app:proof_extra}

For $q=2$, problem \eqref{eq:soft-global} becomes
\begin{equation}
\label{eq:repeat4}
\underset{\substack{\bm{w}^{(1)},\ldots,\bm{w}^{(m)}\in\mathcal{W} \\ \bar{\bm{w}}\in\mathbb{R}^p}}{\min}\,\sum_{k=1}^m\left(\bm{w}^{(k)\top}\bm{\Sigma}^{(k)}\bm{w}^{(k)}+\lambda\bm{w}^{(k)\top}\bm{w}^{(k)}+\gamma(\bar{\bm{w}}-\bm{w}^{(k)})^{\top}(\bar{\bm{w}}-\bm{w}^{(k)})\right).
\end{equation}
The optimal value for the parameter $\bar{\bm{w}}$ can be solved explicitly as it appears only in the last term of \eqref{eq:repeat4}, which is a simple quadratic function. The solution is $\bar{\bm{w}}^\star=m^{-1}\sum_{k=1}^m\bm{w}^{(k)}$.
Next, the last term in \eqref{eq:repeat4} can be further decomposed as
\begin{equation}
\label{eq:quadr}
\sum_{k=1}^m\gamma(\bar{\bm{w}}^\star-\bm{w}^{(k)})^{\top}(\bar{\bm{w}}^\star-\bm{w}^{(k)})=\sum_{k=1}^m\gamma(\bar{\bm{w}}^{\star\top}\bar{\bm{w}}^\star-\bar{\bm{w}}^{\star\top}\bm{w}^{(k)}-\bm{w}^{(k)\top}\bar{\bm{w}}^\star+\bm{w}^{(k)\top}\bm{w}^{(k)}).
\end{equation}
The second term in \eqref{eq:quadr} is equal to the first term (but with the opposite sign) as $\sum_{k=1}^m \bar{\bm{w}}^{\star\top}\bm{w}^{(k)} = \bar{\bm{w}}^{\star\top} \sum_{k=1}^m \bm{w}^{(k)} = m \bar{\bm{w}}^{\star\top}\bar{\bm{w}}^\star = 
\sum_{k=1}^m \bar{\bm{w}}^{\star\top}\bar{\bm{w}}^\star$. The same holds for the third term in \eqref{eq:quadr}. Now our optimisation \eqref{eq:repeat4} can be written as
\begin{equation*}
\label{eq:repeat4_rewrite}
\underset{\bm{w}^{(1)},\ldots,\bm{w}^{(m)}\in\mathcal{W} }{\min}\,\sum_{k=1}^m\left(\bm{w}^{(k)\top}\bm{\Sigma}^{(k)}\bm{w}^{(k)}+\lambda\bm{w}^{(k)\top}\bm{w}^{(k)}+\gamma\bm{w}^{(k)\top}\bm{w}^{(k)}-\gamma\bar{\bm{w}}^{\star\top}\bar{\bm{w}}^\star\right).
\end{equation*}
Finally, we can collect the quadratic terms with $\bm{w}^{(k)}$ as $(\bm{w}^{(k)\top}\bm{\Sigma}^{(k)}\bm{w}^{(k)}+\lambda\bm{w}^{(k)\top}\bm{w}^{(k)} + 
\gamma \bm{w}^{(k)\top}\bm{w}^{(k)})=\bm{w}^{(k)\top} [\bm{\Sigma}^{(k)}+(\lambda+\gamma) \bm{I}]\bm{w}^{(k)}$ and we have
\begin{equation*}
\underset{\bm{w}^{(1)},\ldots,\bm{w}^{(m)}\in\mathcal{W} }{\min}\,\sum_{k=1}^m\left(\bm{w}^{(k)\top}[\bm{\Sigma}^{(k)}+(\lambda+\gamma)\bm{I}]\bm{w}^{(k)}-\gamma\bar{\bm{w}}^{\star\top}\bar{\bm{w}}^\star\right),
\end{equation*}
i.e., we have rewritten problem \eqref{eq:soft-global} as problem \eqref{eq:min_wk}.

\subsection{Proof of Proposition~\ref{prop:01}}
\label{app:proof}

Consider the Lagrangian of problem \eqref{eq:min_wk}:
\begin{equation*}
\mathcal{L}=\sum_{k=1}^m\left(\bm{w}^{(k)\top}[\bm{\Sigma}^{(k)}+(\lambda+\gamma)\bm{I}]\bm{w}^{(k)}-\gamma\bar{\bm{w}}^{\star\top}\bar{\bm{w}}^\star\right)+\sum_{k=1}^m\nu^{(k)}(1-\bm{w}^{(k)\top}\bm{1}),
\end{equation*}
where $\nu^{(k)}$ are the Lagrange multipliers for the restrictions on each weight vector $\bm{w}^{(k)}$ that define $\mathcal{W}^\mathrm{opt}$. The first-order conditions are
\begin{equation*}
\frac{\partial\mathcal{L}}{\partial\bm{w}^{(k)}}=2[\bm{\Sigma}^{(k)}+(\lambda+\gamma)\bm{I}]\bm{w}^{(k)}-2\gamma\bar{\bm{w}}^\star-\nu^{(k)}\bm{1}=\bm{0},
\end{equation*}
so 
\begin{equation} \label{eq:app_w_k}
\bm{w}^{(k)}=[\bm{\Sigma}^{(k)}+(\lambda+\gamma)\bm{I}]^{-1}\left(\frac{\nu^{(k)}}{2}\bm{1}+\gamma\bar{\bm{w}}^\star\right).
\end{equation}
By summing up over $k$, we have
\begin{equation*}
m\bar{\bm{w}}^\star=\frac{1}{2}\sum_{k=1}^m\nu^{(k)}[\bm{\Sigma}^{(k)}+(\lambda+\gamma)\bm{I}]^{-1}\bm{1}+\gamma\left\{\sum_{k=1}^m [\bm{\Sigma}^{(k)}+(\lambda+\gamma)\bm{I}]^{-1}\right\}\bar{\bm{w}}^\star,
\end{equation*}
which gives us the average weight 
\begin{equation*}
\bar{\bm{w}}^\star=\frac{1}{2m}\left\{\bm{I}-\frac{\gamma}{m}\sum_{l=1}^{m}[\bm{\Sigma}^{(l)}+(\lambda+\gamma)\bm{I}]^{-1}\right\}^{-1}\left\{\sum_{l=1}^{m} \nu^{(l)}[\bm{\Sigma}^{(l)}+(\lambda+\gamma)\bm{I}]^{-1}\right\}\bm{1},
\end{equation*}
where the summation index is changed from $k$ to $l$. We can now substitute the average weight in \eqref{eq:app_w_k}, so that the optimal solution must satisfy
\begin{equation}
\label{eq:app_w_k_long}
\begin{split}
\bm{w}^{(k)}=\frac{\nu^{(k)}}{2}[\bm{\Sigma}^{(k)}+(\lambda+\gamma)\bm{I}]^{-1}\bm{1}+\frac{\gamma}{2m}[\bm{\Sigma}^{(k)}+(\lambda+\gamma)\bm{I}]^{-1}&\left\{\bm{I}-\frac{\gamma}{m}\sum_{l=1}^{m}[\bm{\Sigma}^{(l)}+(\lambda+\gamma)\bm{I}]^{-1}\right\}^{-1} \\
&\qquad\times\left\{\sum_{l=1}^{m} \nu^{(l)}[\bm{\Sigma}^{(l)}+(\lambda+\gamma)\bm{I}]^{-1}\right\}\bm{1}.
\end{split}
\end{equation}
We have $m$ restrictions $\bm{w}^{(k)\top}\bm{1}=1$ to identify $\nu^{(k)}$. Let the matrices $A^{(k)}=[\bm{\Sigma}^{(k)}+(\lambda+\gamma)\bm{I}]^{-1}$ and $B=\left(\bm{I}-\gamma/m\sum_{l=1}^{m} A^{(l)}\right)^{-1}$, so the expression \eqref{eq:app_w_k_long} can be shortened to 
\begin{equation}
\label{eq:app_w_k_short}
\bm{w}^{(k)}=\frac{\nu^{(k)}}{2} A^{(k)}\bm{1}+\frac{\gamma}{m}A^{(k)}B\left(\sum_{l=1}^{m}\frac{\nu^{(l)}}{2}A^{(l)}\right)\bm{1}.
\end{equation}
The above expression allows us to compute the average weight vector as
\begin{equation*}
\bar{\bm{w}}^\star=\frac{1}{m}\sum_{k=1}^m\frac{\nu^{(k)}}{2}A^{(k)}\bm{1}+\frac{\gamma}{m^2}\sum_{k=1}^mA^{(k)}B\left(\sum_{l=1}^{m}\frac{\nu^{(l)}}{2}A^{(l)}\bm{1}\right),
\end{equation*}
and 
\begin{equation*}
\sum_{k=1}^m\frac{\nu^{(k)}}{2}A^{(k)}\bm{1}=m\left(\bm{I}+\frac{\gamma}{m}\sum_{k=1}^mA^{(k)}B\right)^{-1}\bar{\bm{w}}^\star
\end{equation*}
that can be substituted back in the last term of \eqref{eq:app_w_k_short}. With $D=\left(\bm{I}+\gamma/m\sum_{k=1}^mA^{(k)}B\right)^{-1}$, we now have
\begin{equation*}
\bm{w}^{(k)}=\frac{\nu^{(k)}}{2}A^{(k)}\bm{1}+\gamma A^{(k)}BD\bar{\bm{w}}^\star.
\end{equation*}
Using $\bm{w}^{(k)\top}\bm{1}=1$, we can find
\begin{equation*}
 \frac{\nu^{(k)}}{2}=\frac{1-\gamma\bm{1}^\top A^{(k)}BD\bar{\bm{w}}^\star}{\bm{1}^\top A^{(k)}\bm{1}}
\end{equation*}
and express the individual weights as
\begin{equation*}
\bm{w}^{(k)}=\frac{1-\gamma\bm{1}^\top A^{(k)}BD\bar{\bm{w}}^\star}{\bm{1}^\top A^{(k)} \bm{1}}A^{(k)}\bm{1}+\gamma A^{(k)}BD\bar{\bm{w}}^\star.
\end{equation*}

\section{Synthetic data experiments}
\label{app:B}

\subsection{Time series visualisation}
\label{app:visualisation}

\begin{figure}[ht!]
\centering
\includegraphics{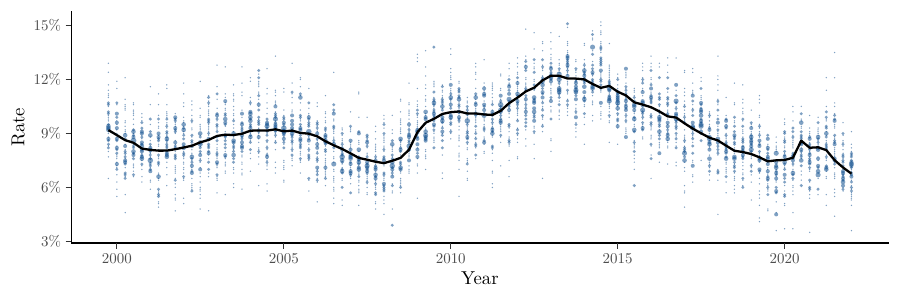}
\caption{Synthetic forecasts for unemployment. Points represent forecasts and lines denote actual values of the forecast target. Point sizes reflect the distribution of forecasts, with larger points reflecting more forecasts at that level. To keep the scale of the forecast errors in line with the scale of the unemployment series, we set the minimum standard deviation $a=0.62$, which is the minimum of the real forecast error sample standard deviations, and the maximum standard deviation $b=3a$.}
\label{fig:simulation-time-series}
\end{figure}

We visually illustrate the simulation design of Section~\ref{sec:synthetic}. Recall the forecast errors are generated directly and assumed unbiased. For an interesting visualisation, we add these forecast errors to the unemployment series studied in the SPF exercise to obtain forecasts.\footnote{ The forecast errors could be added to any underlying time series without changing the simulation results since the mean and correlation structure of the forecast errors will remain unchanged.} Figure~\ref{fig:simulation-time-series} plots the synthetic forecasts alongside the actual values of unemployment.
The synthetic forecasts are similar in variation to the real forecasts of Figure~\ref{fig:spf-data}. The only substantial difference in Figure~\ref{fig:spf-data} seems to be some periods of consistent under/over forecasting. Forecast bias is difficult to estimate \citep{Gibbs2021}, so it would be interesting to consider in future work if globalisation can help produce more precise estimates of such bias.

\subsection{Varying sample sizes}
\label{app:larger}

\begin{figure}[p]
\centering
\includegraphics{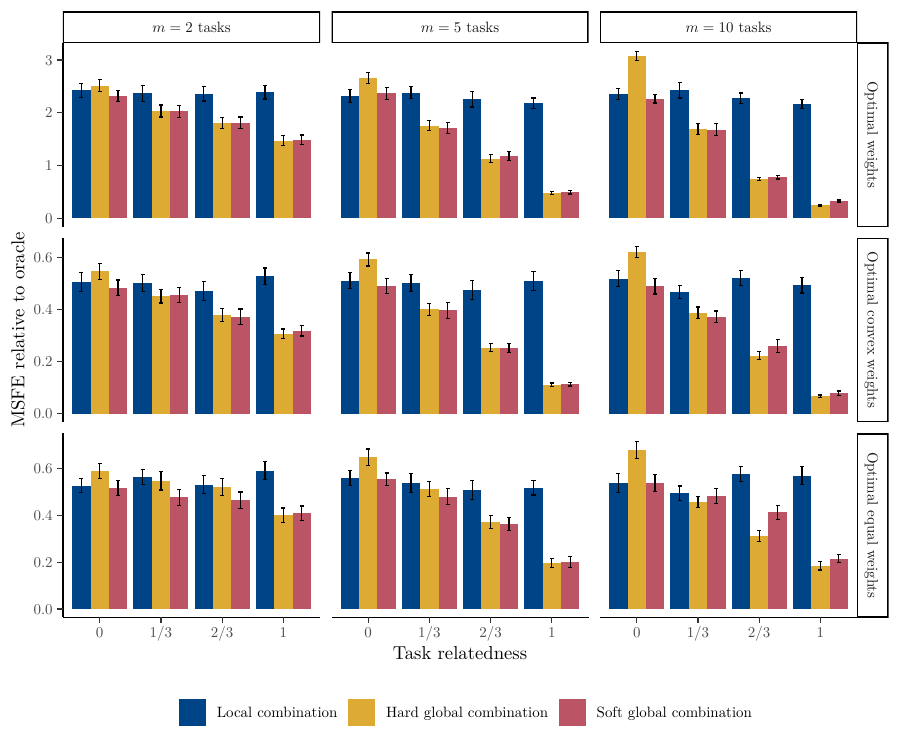}
\caption{Mean square forecast error as a function of task-relatedness parameter $\alpha$ for 30 synthetic datasets with $p=50$ forecasters and $T=25$ samples. Vertical bars represent averages and error bars denote one standard errors. All values are relative to oracle weights.}
\label{fig:simulation-25}
\end{figure}

\begin{figure}[p]
\centering
\includegraphics{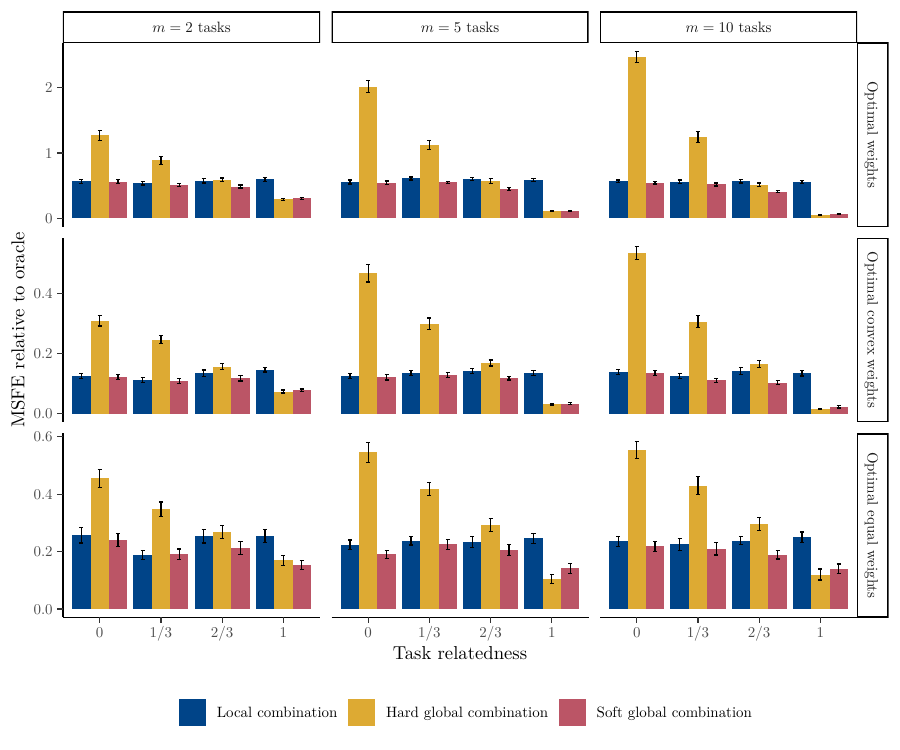}
\caption{Mean square forecast error as a function of task-relatedness parameter $\alpha$ for 30 synthetic datasets with $p=50$ forecasters and $T=100$ samples. Vertical bars represent averages and error bars denote one standard errors. All values are relative to oracle weights.}
\label{fig:simulation-100}
\end{figure}

\begin{figure}[p]
\centering
\includegraphics{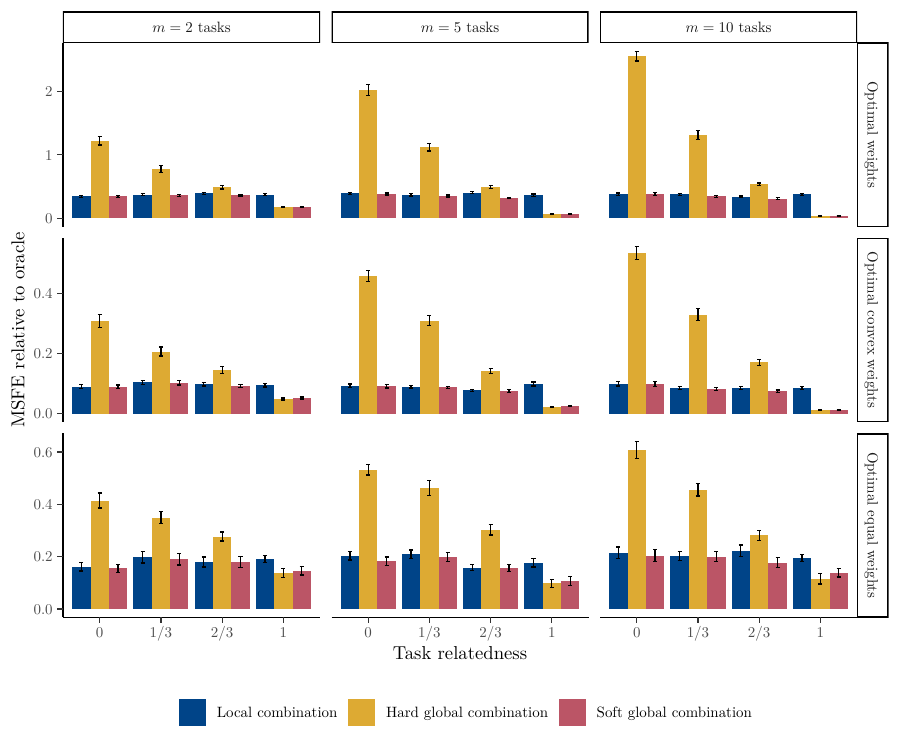}
\caption{Mean square forecast error as a function of task-relatedness parameter $\alpha$ for 30 synthetic datasets with $p=50$ forecasters and $T=150$ samples. Vertical bars represent averages and error bars denote one standard errors. All values are relative to oracle weights.}
\label{fig:simulation-150}
\end{figure}

Figures \ref{fig:simulation-25}, \ref{fig:simulation-100}, and \ref{fig:simulation-150} extend the simulations results of Section~\ref{sec:synthetic} with smaller and larger sample sizes of $T\in\{25,100,150\}$. The number of forecasters fixed at $p=50$.

For a shorter history of $T=25$, Figure~\ref{fig:simulation-25} shows little difference between the local and hard local combinations when tasks are not related and the number of tasks is small ($m=2$ or 5). In other words, even when the additional global information is irrelevant, it does not hurt the performance of hard global combination. The proposed soft global combination performs similarly in this case, even though an extra parameter $\gamma$ is introduced into the problem. Once task-relatedness increases substantially to 2/3 or 1, there is a clear benefit to using global information. While the performance of local combination stagnates, our proposed soft global approach is not harmed by it and is similar to hard global combination. These results suggest using soft global combination for shorter series as it performs well even when the tasks are not related and significantly improves when they are related.

When longer history is available ($T=100$ or 150), Figures \ref{fig:simulation-100} and \ref{fig:simulation-150} show that hard global combination is harmful when task-relatedness is low or moderate and even when it is strong ($\alpha=2/3$) for $T=150$. The reason for this result is the large amount of \emph{irrelevant} information introduced by long time series. The irrelevant information does not harm our proposed soft global combination as $\gamma$ can be tuned to remove its effect. Soft global combination performs similarly to local combination when the task-relatedness is low, moderate, or even strong ($\alpha=2/3$). Of course, when the tasks are perfectly related ($\alpha=1$), the situation flips. Now, hard global combination is highly beneficial as long series bring a lot of \emph{relevant} information. Again, soft global combination can extract the same benefits as hard global combination.

These results confirm the previous suggestion of using soft global combination, but now for longer series, as it can perform well whether tasks are related or not. In practice, one does not usually know the relatedness of the tasks. In our empirical study, e.g., inflation and unemployment are related by the Phillips curve. This relationship changes over time with periods where the variables are strongly related and periods where they are weakly related. The advantage of soft global combination is that one does not need to know the strength of the relation to witness benefits.

\begin{figure}[ht!]
\centering
\includegraphics{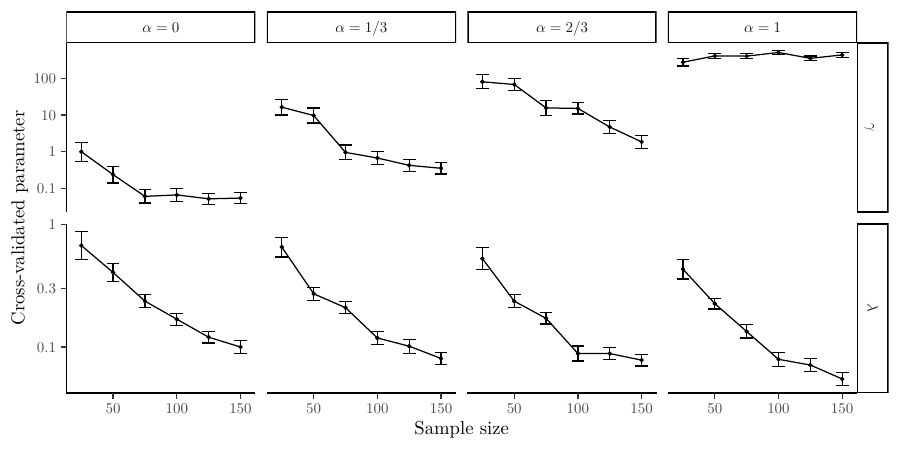}
\caption{Cross-validated globalisation parameter $\gamma$ and shrinkage parameter $\lambda$ as a function of sample size $n$ and task-relatedness $\alpha$ for 100 synthetic datasets with $p=50$ forecasters $m=2$ tasks. Solid points represent averages and error bars denote one standard errors. Optimal weights are used.}
\label{fig:tuning-behavior}
\end{figure}

Figure~\ref{fig:tuning-behavior} shows the behaviour of the cross-validated globalisation parameter $\gamma$ and cross-validated shrinkage parameter $\lambda$.
As the sample size $T$ increases, the globalisation parameter $\gamma$ decreases, giving more weight to local information. By tapering down $\gamma$, soft global combination can deal with the additional noise that is harmful to hard global combination even when the task-relatedness is moderate and especially when it is low. When the tasks are perfectly related, $\gamma$ remains at its maximum. The shrinkage parameter $\lambda$ also decreases with sample size as the covariance matrix estimator becomes increasingly reliable.

\subsection{Penalty comparisons}
\label{app:penalty}

\begin{figure}[p]
\centering
\includegraphics{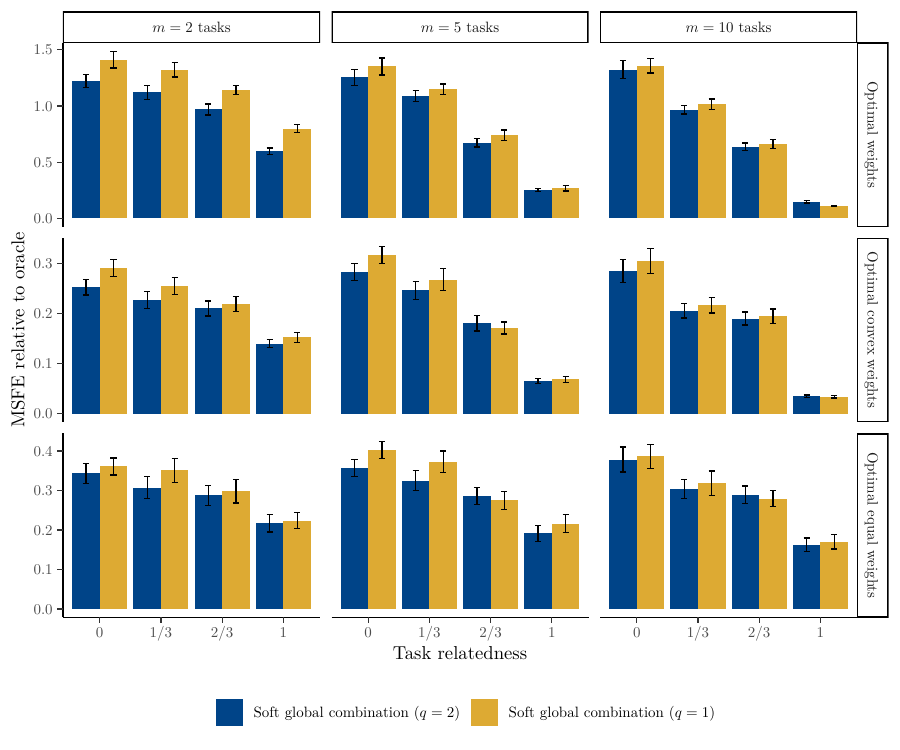}
\caption{Mean square forecast error as a function of task-relatedness parameter $\alpha$ for 30 synthetic datasets with $p=50$ forecasters and $T=100$ samples. Vertical bars represent averages and error bars denote one standard errors. All values are relative to oracle weights.}
\label{fig:simulation-q1q2}
\end{figure}

Figure~\ref{fig:simulation-q1q2} compares soft global combination \eqref{eq:soft-global} configured with the squared deviation globalisation penalty ($q=2$) against the absolute deviation penalty ($q=1$). The simulation design is the same as that in Section~\ref{sec:synthetic}.
The squared deviation penalty often has an edge over the absolute deviation penalty, possibly due to it being continuous. Often, however, the two penalties are within statistical precision of one another. From an optimisation standpoint, the squared deviation penalty is simpler to model since it does not require additional variables and constraints to handle the absolute values. Hence, while our implementation supports both penalties, we default this parameter to $q=2$.

\subsection{Equal weights benchmark}
\label{app:equal}

\begin{figure}[p]
\centering
\includegraphics{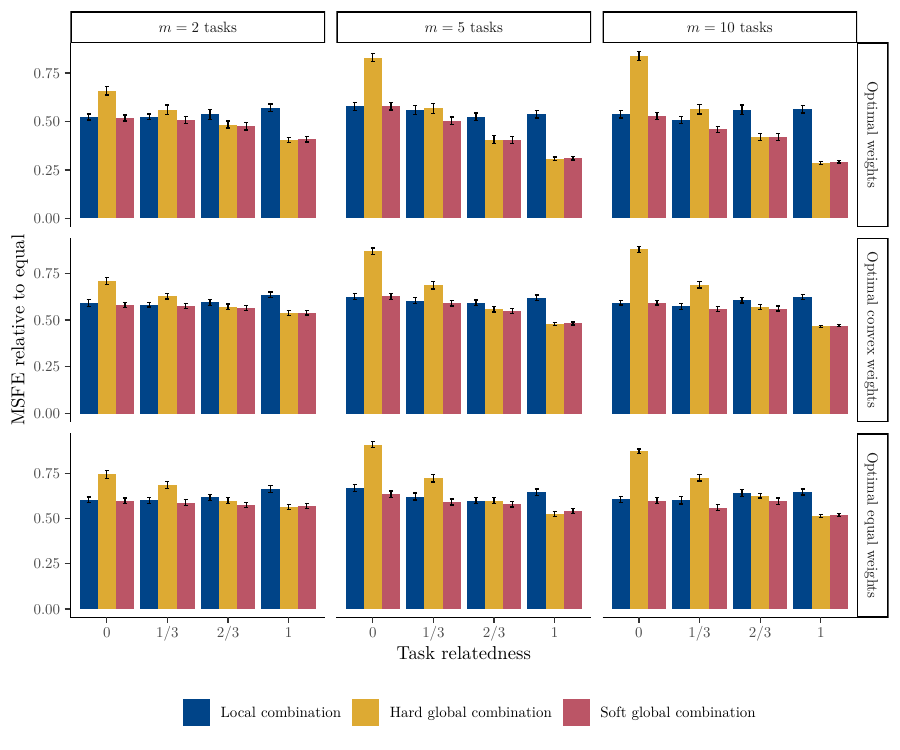}
\caption{Mean square forecast error as a function of task-relatedness parameter $\alpha$ for 30 synthetic datasets with $p=50$ forecasters and $T=50$ samples. Vertical bars represent averages and error bars denote one standard errors. All values are relative to equal weights.}
\label{fig:simulation-equal}
\end{figure}

Figure~\ref{fig:simulation-equal} reports the same results as Figure~\ref{fig:simulation-50} but now stated relative to equal weights rather than oracle weights:
\begin{equation*}
\text{MSFE relative to equal}:=\frac{\hat{\bm{w}}^{(1)\top}\bm{\Sigma}^{(1)}\hat{\bm{w}}^{(1)}}{\bm{1}^\top\bm{\Sigma}^{(1)}\bm{1}p^{-2}}.
\end{equation*}
The denominator is the mean square forecast error from equal weights.
 Because this metric is simply a rescaling of that in Figure~\ref{fig:simulation-equal}, all key observations carry over here. Consistent with the empirical evidence of Section~\ref{sec:tuned}, optimal weights stand to gain the most from globalisation. In short, if task-relatedness is high, or if the number of tasks is large, optimal weights are an excellent choice for soft global combination.

\section{Survey of Professional Forecasters}
\label{app:C}

\subsection{Cross-validation tuning vs. full-information tuning}
\label{app:cv}

To glean further insight into the the operational performance of cross-validation tuning for the globalisation parameter $\gamma$, we provide comparisons with full-information tuning, where $\gamma$ is chosen with knowledge of all future values of the forecast target. Table~\ref{tab:spf-tuning-gamma} reports the results from these comparisons on the SPF over the period 2017 Q1 to 2019 Q4. The first two columns of numbers contain the per-task mean square forecast errors. The second two columns contain the corresponding average of the (log) globalisation parameter with standard errors in parentheses.
\begin{table}[ht!]
\centering
\footnotesize
\begin{tabularx}{390pt}{lRRRR}
\toprule
 & \multicolumn{2}{r}{MSFE relative to equal} & \multicolumn{2}{r}{Avg. globalisation param. $(\log\gamma)$} \\ 
\cmidrule{2-3} \cmidrule(l){4-5} & Full-info. & Cross-valid. & Full-info. & Cross-valid. \\ 
\midrule
\multicolumn{4}{l}{\emph{Optimal weights}} \\ 
One-year growth & 0.716 & 0.741 & 6.908 & -3.966 (0.516) \\ 
Two-year growth & 0.589 & 0.657 & 2.303 & -1.663 (1.302) \\ 
One-year inflation & 1.125 & 1.206 & 6.908 & 3.198 (0.812) \\ 
Two-year inflation & 1.669 & 1.720 & 6.908 & 5.501 (1.060) \\ 
One-year unemployment & 0.360 & 0.360 & -6.908 & -5.373 (0.000) \\ 
Two-year unemployment & 0.272 & 0.451 & -6.908 & 0.895 (0.790) \\ 
\multicolumn{5}{l}{\emph{Optimal convex weights}} \\ 
One-year growth & 0.876 & 0.867 & 0.768 & -1.919 (0.571) \\ 
Two-year growth & 0.944 & 0.957 & 0.768 & 6.268 (0.640) \\ 
One-year inflation & 1.075 & 1.104 & 6.908 & -1.407 (1.077) \\ 
Two-year inflation & 1.099 & 1.183 & -0.768 & 3.710 (0.875) \\ 
One-year unemployment & 0.893 & 0.894 & -6.908 & -4.861 (0.577) \\ 
Two-year unemployment & 0.873 & 0.878 & -2.303 & -5.373 (0.598) \\ 
\multicolumn{5}{l}{\emph{Optimal equal weights}} \\ 
One-year growth & 0.861 & 0.850 & -6.908 & -0.640 (0.693) \\ 
Two-year growth & 0.967 & 0.967 & 2.303 & 4.605 (0.694) \\ 
One-year inflation & 1.080 & 1.008 & -0.768 & -0.768 (1.165) \\ 
Two-year inflation & 1.049 & 1.146 & -0.768 & 2.047 (0.798) \\ 
One-year unemployment & 0.885 & 0.923 & 2.303 & -2.558 (0.172) \\ 
Two-year unemployment & 0.882 & 0.936 & 0.768 & -3.198 (0.550) \\ 
\bottomrule
\end{tabularx}

\caption{Comparisons of cross-validation tuning with full-information tuning for the Survey of Professional Forecasters over the period 2017 Q1 to 2019 Q4. The globalisation parameter $\gamma$ is reported in logarithmic scale. The shrinkage parameter $\lambda=0.1$. All tasks are grouped. Standard errors are reported in parentheses.}
\label{tab:spf-tuning-gamma}
\end{table}
The MSFEs from cross-validation are, on average, 6\% larger than the full-information MSFEs. The $\gamma$ from full-information tuning varies between the minimum and maximum of the parameter grid ($\log0.001\approx-6.908$ and $\log1000\approx6.908$, respectively), reinforcing the need for soft global combination. The $\gamma$ chosen via cross-validation is sometimes close to that of the full-information procedure and other times further away. This variation is likely due to estimation error from cross-validation, which is potentially considerable given the relatively small training sample. Structural breaks may also play a role and could be addressed via cross-validation with a rolling window. Variability in the cross-validated value of $\gamma$ is typically largest for the two-year horizon tasks, likely because these are more difficult than the one-year tasks and hence noisier. Though cross-validation is a practical tool for tuning $\gamma$, there is certainly value in future work investigating alternatives (e.g., a formulaic characterisation of the optimal $\gamma$ via asymptotic analysis).

\begin{figure}[ht!]
\centering
\includegraphics{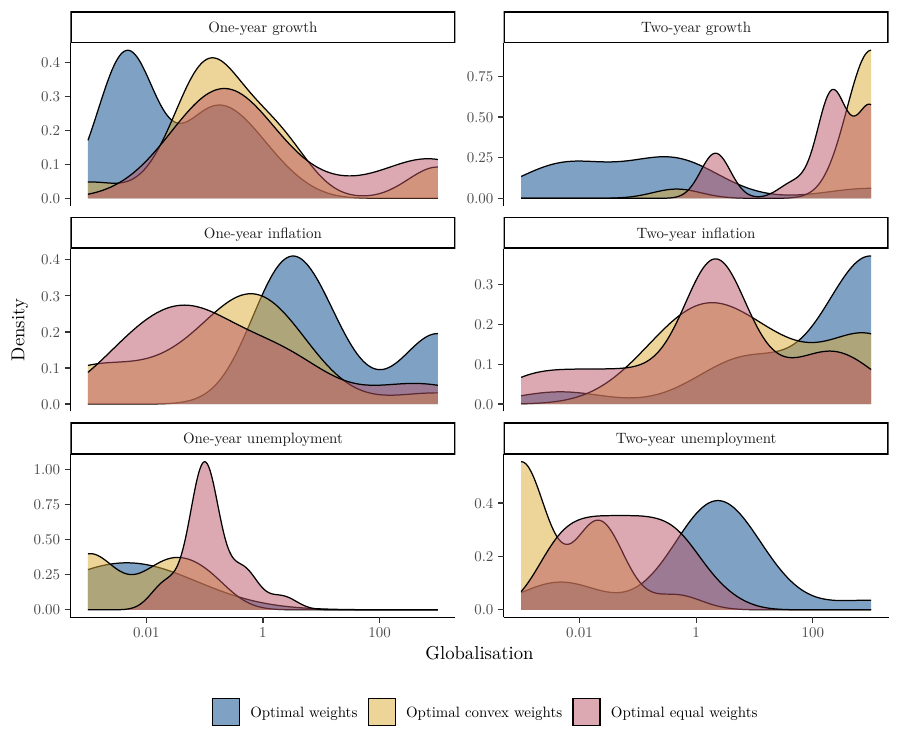}
\caption{Cross-validated globalisation parameter $\gamma$ for the Survey of Professional Forecasters over the period 2017 Q1 to 2019 Q4. All tasks are grouped. The shrinkage parameter $\lambda=0.1$. The $x$-axis is in log scale.}
\label{fig:spf-tuning-gamma}
\end{figure}

For further insight into the globalisation parameters from by cross-validation, we present Figure~\ref{fig:spf-tuning-gamma} which provides kernel density estimates for the distribution of the cross-validated $\gamma$ (in log scale).
The highest-density regions are typically on the interior of the support. This characteristic is always present for optimal equal weights and is consistent with Figure~\ref{fig:spf-path-optimal-equal} where the best $\gamma$ is always some intermediate value. For the other (continuous) weighting schemes, the distributions sometimes have highest density at the extreme points (fully global or fully local). For example, the distributions of optimal weights with two-year inflation and optimal convex weights with two-year growth have their highest modes at the largest value of $\gamma$. These cases are consistent with the minima along the globalisation paths in Figures~\ref{fig:spf-path-optimal} and \ref{fig:spf-path-optimal-convex}.

\section{M4 Competition}
\label{app:m4}

This section complements the paper's main results with additional experiments on time series from the M4 Competition \citep{Makridakis2020}. M4 involved 100,000 forecasting tasks spanning numerous domains and forecast horizons. Given this paper's focus on low-frequency, economic-orientated time series, we focus on a subset of this data that has adequate historic data available:
\begin{itemize}
\item All quarterly economic time series beginning in 1960 Q1 and ending in 2007 Q5 (70 series);
\item All monthly economic time series beginning in Jan 1996 and ending in May 2008 (30 series); and
\item All weekly finance time series beginning in week 5 2000 and ending in week 24 2009 (16 series).
\end{itemize}
Each collection of time series above constitutes a group of related forecast tasks. These experiments thus extend those in the main paper by (a) considering more tasks (up to 70 tasks) and (b) considering different time series frequencies (quarterly, monthly, and weekly).

To generate forecasts for the individual time series, we follow \citet{Montero-Manso2020} and use nine models implemented in the \texttt{R} package \texttt{forecast} \citep{Hyndman2023}. These models include a random walk, autoregressive integrated moving average, and exponential smoothing, among others \citep[see][]{Montero-Manso2020}. We take the first 40\% of each time series and fit each model to these initial training sets. The following observation is forecast ($h=1$) before it is added to the training set. This process leads to one-step ahead forecast errors for the last 60\% of each time series.\footnote{The forecast errors for each series are made scale-free by dividing by the root mean square forecast error from a seasonal random walk taken over the training set.} We then evaluate our global combinations by splitting the forecast errors into training, validation, and testing sets in 0.4-0.4-0.2 proportions with the temporal ordering preserved. In fitting the combination weights to the training set, \texttt{Gurobi} is given a 300 second time limit on AMD Ryzen Threadripper 3970x.

\begin{figure}[ht!]
\centering
\includegraphics{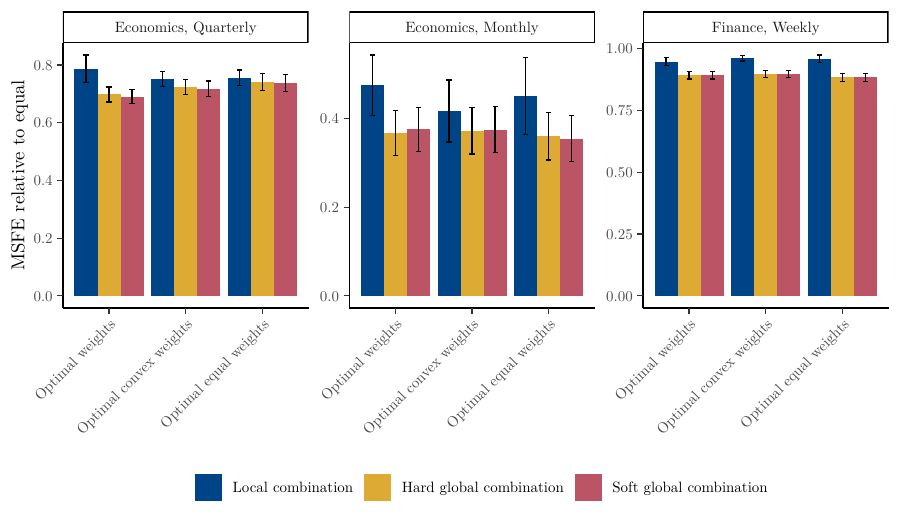}
\caption{Mean square forecast error for economic and financial data from the M4 Competition. Vertical bars represent averages and error bars denote one standard errors. All values are relative to equal weights.}
\label{fig:m4}
\end{figure}

Figure~\ref{fig:m4} reports mean square forecast errors (relative to equal weights) on the testing sets with any tuning parameters chosen on the validation sets.
The benefits from globalisation are most significant for the monthly series, though the other two frequencies witness performance improvements too. Interestingly, there is no distinct difference between hard and soft global combinations; their performance is usually nearly identical. This result suggests the similarity between the tasks is high. Fortunately, soft global combination well adapts to this similarity across all weighting schemes. In contrast to the Survey of Professional Forecasters exercise in Section~\ref{sec:survey}, equal weights are not a strong contender here. Upon closer inspection, their poorer performance seems due to a few bad models that are otherwise zeroed out by the data-driven combinations. \citet{Montero-Manso2020} observe a similar phenomenon where a small subset of the nine models dominate their forecast combinations most of the time.

\end{document}